\documentclass[%
 reprint,
 amsmath,amssymb,
 aps,
]{revtex4-2}

\usepackage{graphicx}
\usepackage{dcolumn}
\usepackage{bm,ulem,color}

\begin{document}

\title{Surface acoustic wave-driven valley current generation in intervalley coherent states}

\author{Hiroto Tanaka}
\email[]{tanaka.hiroto.54z@st.kyoto-u.ac.jp}
\affiliation{%
Department of Physics, Graduate School of Science, Kyoto University, Kyoto 606-8502, Japan
}%

\author{Youichi Yanase}
\affiliation{%
Department of Physics, Graduate School of Science, Kyoto University, Kyoto 606-8502, Japan
}%

\begin{abstract}
Recent experiments have reported valley-gauge-symmetry-broken phases, identified as intervalley coherent (IVC) states. 
Exploration of anomalous responses, particularly those analogous to superconductivity, has become an urgent theoretical issue.  
In this study, we show that the IVC order gives rise to anomalous valley-current generation driven by surface acoustic waves (SAWs). 
The anomalous valley current exhibits a characteristic power-law dependence for low-frequency SAWs. 
Furthermore, we demonstrate by numerical analysis that the IVC order significantly enhances valley-current generation in rhombohedral graphene. 
These results open a pathway toward exploring exotic phenomena emerging from valley-gauge-symmetry breaking, in close analogy with gauge-symmetry breaking in superconductors.
\end{abstract}

\date{\today}

\maketitle

\emph{Introduction.---}
Various types of spin and valley ordering have recently been reported in two-dimensional materials with the honeycomb structure. 
Notably, the intervalley coherent (IVC) state has been focused as an exotic ordered state that spontaneously breaks valley gauge symmetry~\cite{Liu2022, Farahi2023, Fan2025, Liao2025, Liu2024, Dumitru2022, Hong2022, Coissard2022, Nuckolls2023, Kim2023, Po2018, Kwan2021, Bultinck2020, Lian2021, Zhang2020, Vituri2025, Chatterjee2022, Das2024, Patri2023, Arp2024, Munoz2025, Das2025, Wei2025, Thomson2022}.
The analogy to superconductivity, which breaks gauge symmetry, implies the appearance of characteristic transport and response phenomena emerging from valley-gauge-symmetry breaking.
However, only a few theoretical works have proposed the transport signature of the IVC state~\cite{Das2025, Wei2025, Thomson2022}, and the IVC state has been detected mainly through scanning tunneling microscopy (STM) and spectroscopy (STS) experiments~\cite{Nuckolls2023, Kim2023,Coissard2022, Liu2022, Farahi2023, Fan2025, Liao2025, Liu2024,Dumitru2022, Hong2022}.

Surface acoustic waves (SAWs) are mechanical vibrations that propagate within the unbounded surface of the elastic media~\cite{Nie2023}. 
When the target two-dimensional materials are placed on the elastic medium, the SAWs induce spatio-temporal modulation of the target materials.
Specifically in honeycomb systems, the spatial modulation of hopping integrals can be incorporated as a pseudogauge field. 
The pseudogauge field affects the dynamics of the electrons~\cite{von_Oppen2009,Vaezi2013, Sela2020, Zhao2022, Bhalla2022, Cazalilla2014, Sukhachov2020, Matsumoto2025}.
The pseudogauge field is similar to the valley gauge field, which couples to the electrons with opposite charge between $K$ and $K^{\prime}$ points in momentum space in contrast to the standard electromagnetic field. 
Thus, the pseudogauge field activates the valley degree of freedom and can drive even charge-neutral quasiparticles~\cite{Uchoa2013, Massarelli2017, Nica2018, Nayga2019, Sano2024}.

\begin{figure}[tbp]
 \includegraphics[width=1.0\linewidth]{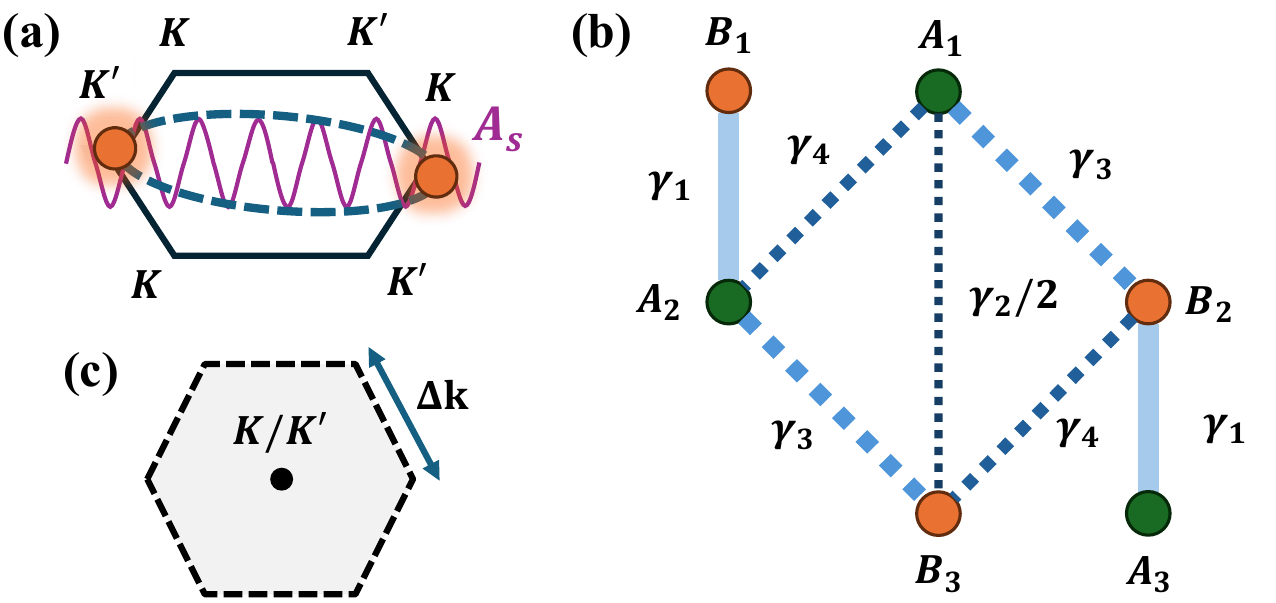}
 \caption{(a) Schematic of the IVC state under the SAWs. The SAWs induce the pseudogauge field $\bm{A}_{s}$. (b) The unit cell and the interlayer hopping process in R3G. Here, $A_j$ and $B_j$ represent the sublattice on the layer $j=1,2,3$. The hopping integrals are denoted by $\gamma_i$.  (c) The valley charge is assumed to be well-defined in the shaded domain $D$ around the K/K' points.  
 We parameterize the size of the domain $D$ by $\Delta k$.}
 \label{fig:shematic}
\end{figure}

Here, we explore a pseudogauge-field-driven response phenomenon arising from spontaneous valley-gauge-symmetry breaking in the IVC state [Fig.~\ref{fig:shematic}(a)]. 
The IVC order is a counterpart of superconductivity because the valley ($U(1)$-) gauge symmetry is broken by the former (latter).
The relation between the electromagnetic field and the $U(1)$-gauge symmetry is essential for anomalous electromagnetic responses in superconductors, such as the Meissner effect~\cite{schrieffer2018}. 
We focus on the fact that a similar relation holds between the pseudogauge field and the valley gauge symmetry in the IVC state.

In this Letter, we demonstrate that the IVC order gives rise to anomalous contributions to nonlinear valley currents driven by the pseudogauge field. 
The anomalous valley current generation is associated with the pseudo-superfluid density, which is the derivative of the free energy with respect to the pseudogauge field. 
This is analogous to how the superfluid density determines the electromagnetic properties of superconductors.

We perform numerical calculations of SAW-driven valley-current generation in the tight-binding model of rhombohedral graphene to demonstrate the impact of IVC ordering.
The numerical results show that the frequency dependence follows a characteristic power law, which arises from the valley-gauge-symmetry breaking. 
The IVC order induces a large pseudo-superfluid density and significantly enhances the SAW-driven valley currents.

\emph{Valley gauge field and pseudogauge field.---}
First, we briefly review the similarity between the valley gauge field and the pseudogauge field in monolayer graphene (1-LG)~\cite{von_Oppen2009,Vaezi2013,  Sela2020, Zhao2022, Bhalla2022}, before considering multilayer graphene.
The 1-LG 
is described by the effective Dirac Hamiltonian $H_{\xi}^{\mathrm{1-LG}}(\bm{k})$ per valley $(K/K^{\prime})$ and spin.
The valley gauge field $\bm{A}_{\mathrm{v}}$ is incorporated into the Hamiltonian as $H^{\mathrm{1-LG}}_{\xi}(\bm{k})\rightarrow H^{\mathrm{1-LG}}_{\xi}(\bm{k}+\frac{e}{\hbar}\xi\bm{A}_{\mathrm{v}})$, where $\xi=\pm$ corresponds to the valleys $K$ and $K^{\prime}$.
The valley gauge field, which is a counterpart of the electromagnetic field, is coupled to electrons with opposite signs between the different valleys.
Valley current operators are obtained as derivatives of the Hamiltonian with respect to the valley gauge field~\cite{supplement}.
The pseudogauge fields arise from the modulation of the hopping integrals. 
Mechanical vibrations, such as SAWs, can modulate the hopping integrals and perturb the Hamiltonian as
$H^{\mathrm{1-LG}}_{\xi}(\bm{k})\rightarrow H^{\mathrm{1-LG}}_{\xi}(\bm{k}+\frac{e}{\hbar}\xi\bm{A}_{\mathrm{s}})$, where $\bm{A}_{\mathrm{s}}$ is the pseudogauge field.
Because the two fields are formally equivalent, the pseudogauge field $\bm{A}_{\mathrm{s}}$ can be interpreted as an external valley gauge field.
Thus, anomalous responses driven by the pseudogauge field are expected to emerge from spontaneous valley-gauge-symmetry breaking in the IVC state.

In multilayer graphene, the couplings of the pseudogauge field and electrons deviate from those of the valley gauge field, but a similar relation exists.
We later discuss the details of the differences in the couplings.

\emph{Rhombohedral graphene.---}
We consider rhombohedral trilayer graphene (R3G) under the displacement field, which is one of the promising platforms for the IVC states~\cite{Chatterjee2022, Vituri2025, Liu2024}. 
R3G in the normal state is accurately described using a six-band model per valley $(K/K^{\prime})$ and spin~\cite{Zhang2010, Jung2013}. 
The interlayer hopping processes are illustrated in Fig.~\ref{fig:shematic}(b).
The interlayer hopping parameters are defined as $\gamma_i$ ($i=1-4$), while the intralayer nearest-neighbor hopping parameter is $\gamma_{0}$.
All our numerical calculations are performed for the six-band model, with tight-binding parameters taken from Ref.~\cite{Zhou2021}. 
The six-band model Hamiltonian $H_{\xi}(\bm{k})$ is written in the basis $(A_{1}, B_{3}, B_{1}, A_{2}, B_{2}, A_{3})$.
The matrix elements are given by $\pi=\hbar(\xi k_{x}+ik_{y})$ [$\pi^{\dagger} = \hbar(\xi k_{x}-ik_{y})$] and some constant parameters~\cite{supplement}.
The point group symmetry of the model is $\mathrm{C_{3v}}$ under the displacement field.
We assume that the valley charge is well-defined in the $k$-space domain $D$ around the valley points [Fig.~\ref{fig:shematic}(c)].

\begin{table}[tbp]
    \centering
    \begin{tabular}{c|c|c|c}
    type & $\mathcal{H}^{\alpha\beta}_{\xi}(\bm{k})$ & $(\alpha, \beta)$ &  layer \\ \hline
    A & $v_{0}\pi$ &$(A_{l}, B_{l})\quad (l=1, 2,3)$ & intra-layer\\
    B & const. & $(B_{1}, A_{2}), (B_{2}, A_{3}), (A_{1}, B_{3})$ & inter-layer\\
    C1 & $v_{3}\pi$ & $(B_{2}, A_{1}), (B_{3}, A_{2})$  & inter-layer\\
    C2 & $v_{4}\pi$ & $(A_{1}, A_{2}), (B_{2}, B_{3})$  & inter-layer
    \end{tabular}
    \caption{Off-diagonal matrix elements of the six-band model Hamiltonian $\mathcal{H}^{\alpha\beta}_{\xi}(\bm{k})$ are classified into the types A, B, C1, and C2 based on the coupling to the pseudogauge field. The type B matrix elements are not coupled to the pseudogauge field, since they do not depend on the wave vector $\bm{k}$.}
    \label{tab:placeholder}
\end{table}

To describe the IVC states, we construct the mean-field Hamiltonian,
\begin{align}
  \mathcal{H}_{\mathrm{MF}} = \sum_{\bm{k}, \alpha, \beta}
  \psi_{\alpha}^{\dagger}(\bm{k})
  \begin{pmatrix}
    H^{\alpha\beta}_{+}(\bm{k}) & \Delta^{\alpha\beta} \\
    \Delta^{\dagger\alpha\beta} & H^{\alpha\beta}_{-}(\bm{k})
  \end{pmatrix}
    \psi_{\beta}(\bm{k}), \label{eq:MF_Hamiltonian}
\end{align}
where we introduce the fermion operator $\psi_{\alpha}(\bm{k})=(c_{\alpha}(\bm{K}+\bm{k}), c_{\alpha}(\bm{K}^{\prime}+\bm{k}))^{\top}$, and the index $\alpha$ represents the spin, layer, and sublattice degrees of freedom. 
We assume a $\mathcal{T}$-symmetric IVC order which preserves the point group symmetry $\mathrm{C_{3v}}$~\cite{supplement}.

\emph{SAWs and pseudogauge field.---}
Rayleigh-type SAWs propagating on the surface of piezoelectric substrates in the $xy$ plane are described by the displacement field $\bm{u}(\bm{r},t)=\mathrm{Re}\left[\left(u_{L}\hat{Q}+iu_{z}\hat{z}\right)e^{i(\bm{Q}\cdot\bm{r}-\Omega t)}\right]$, where $u_{L}$ and $u_{z}$ are the longitudinal and out-of-plane components~\cite{Mandal2022}. 
The SAWs propagate with the phonon wave vector $\bm{Q}=Q(\cos\phi, \sin\phi)$ and frequency $\Omega$.

The SAWs generate the pseudogauge field via modulation of the hopping integrals.  
In monolayer graphene, where only the intralayer hopping processes occur, the pseudogauge field coincides with the valley gauge field. 
On the other hand, in the R3G, interlayer hopping processes are significant for electronic properties such as energy bands~\cite{Zhang2010, Jung2013},
and it is necessary to consider how the pseudogauge field couples to electrons in these processes.

We derive the pseudogauge field and its coupling to the Hamiltonian from the perturbations of the tight-binding model for the R3G under the lattice modulation $\bm{u}(\bm{r},t)$~\cite{supplement}.
The pseudogauge field couples to electrons like the minimal coupling, but the coupling constants depend on the hopping process.
To summarize the couplings of the pseudogauge field,
we categorize the matrix elements into four types: A, B, C1, and C2, corresponding to the different hopping processes
[Table~\ref{tab:placeholder}].
Type A originates from intralayer hopping, while the other types correspond to interlayer hopping.
The pseudogauge field $\bm{A}_{\mathrm{s}}$ is coupled to the type A elements of Hamiltonian by $H^{\alpha\beta}_{\xi}(\bm{k})\rightarrow H^{\alpha\beta}_{\xi}(\bm{k}+\frac{e}{\hbar}\xi\bm{A}_{\mathrm{s}})$.
The SAW-induced pseudogauge field reads $\bm{A}_{\mathrm{s}}=-\frac{\sqrt{3}\beta_{0}}{2a} u_{L}Q(-\cos 2\phi, \sin 2\phi)\mathrm{Im}\left[e^{i(\bm{Q}\cdot\bm{r}-\Omega t)}\right]$.
Factors $\beta_{i}$ including $\beta_{0}$ are 
defined by $\beta_{i}=-d\log \gamma_{i}/d\log \tau_{i}$, where $\tau_{i}$ is the distance of the hopping process for $\gamma_{i}$.

In type C1 and C2 matrix elements, the pseudogauge field $\bm{A}_{\mathrm{s}}$ couples to electrons with factors $r_{i}$: $H^{\alpha\beta}_{\xi}(\bm{k})\rightarrow H^{\alpha\beta}_{\xi}(\bm{k}+r_{i}\frac{e}{\hbar}\xi\bm{A}_{\mathrm{s}})$, where $i=3,4$ for types C1 and C2, respectively.
The factor $r_{i}$ is determined by $\beta_{i}$ and the hopping distances and directions~\cite{supplement}.
Although the coupling with the pseudogauge field deviates from that with the valley gauge field when $r_{i}\neq 1$, it remains similar to the minimal coupling form.
The out-of-plane component of SAWs also induces 
additional pseudogauge field, $\tilde{\bm{A}}_{\mathrm{s}} \propto O(u_{z})$, in type C1 and C2 matrix elements.
However, we ignore 
$\tilde{\bm{A}}_{\mathrm{s}}$ in the main text because its effect on the valley current generation is negligible~\cite{supplement}.
The effect of SAWs on the type B matrix element is ignorable because the Rayleigh-type SAW does not change the distance of the type B hopping process.

The factors $r_{3}, r_4$ characterize the deviation of the pseudogauge field from the valley gauge field.
In this Letter, we investigate the dependence of the valley current generation on the factors $r_{3}, r_4$ to verify that anomalous contributions originating from the IVC order exist at any $r_{i}$.
In the following, we assume $r_{3}$ is equal to $r_{4}$ for simplicity and introduce the parameter $r=r_{3}=r_{4}$.
The value of $r$ is roughly estimated as $r\sim0.2$~\cite{supplement}.
We calculate the valley current generation with  $ 0 \leq r \leq 1.2$ and verify the robustness of the anomalous valley current generation in the IVC states. 

\emph{Valley acoustogalvanic effect.---}
The oscillating pseudogauge field can drive direct electric currents via nonlinear response mechanisms, known as the acoustogalvanic effect~\cite{Bhalla2022, Sukhachov2020}. 
In this Letter, we focus on the generation of spatio-temporally uniform valley currents driven by a pseudogauge field, which we refer to as the valley acoustogalvanic (VAG) effect.
The stationary valley current can induce valley accumulation, which is advantageous for experimental detection~\cite{Lee2016, Lee2017}. 
Taking account of the pseudogauge field perturbatively, the leading order of the valley current is given by the second-order nonlinear response,
\begin{align}
\left<\mathcal{J}^{\alpha}_{\mathrm{v}}\right>=\sigma^{\alpha;\beta\gamma}_{\mathrm{s},\bm{Q}, -\bm{Q}}(0;\Omega, -\Omega)E^{\beta}_{\mathrm{s}}(\bm{Q}, \Omega)E^{\gamma}_{\mathrm{s}}(-\bm{Q}, -\Omega), \label{eq:VAGE_sigma}
\end{align}
where we introduce the Fourier components of the pseudo-electric field $\bm{E}_{\mathrm{s}}(\bm{r}, t) = -\partial_{t}\bm{A}_{\mathrm{s}}(\bm{r}, t)$.
This formula clarifies the similarity to the nonlinear electric (NLE) conductivity,
\begin{align}
\left<\mathcal{J}^{\alpha}_{\mathrm{e}}\right>=\sigma^{\alpha;\beta\gamma}_{\mathrm{e},\bm{Q}, -\bm{Q}}(0;\Omega, -\Omega)E^{\beta}(\bm{Q}, \Omega)E^{\gamma}(-\bm{Q}, -\Omega). \label{eq:NLE_sigma}
\end{align}
In the following, we assume a sufficiently large wavelength of the SAW and set $\bm{Q}\simeq \bm{0}$.

We consider the VAG conductivity $\sigma^{\alpha;\beta\gamma}_{\mathrm{s}} = \sigma^{\alpha;\beta\gamma}_{\mathrm{s},\bm{0}, -\bm{0}}$ in the regime $\Omega\ll \tau^{-1}_{\mathrm{rel}}\ll \Delta E$, where $\tau_{\mathrm{rel}}$ is the relaxation time and $\Delta E$ is the energy difference of the electron bands. 
We obtain the analytical formula of the VAG conductivity by generalizing the theoretical work of the NLE conductivity~\cite{Watanabe2022, Watanabe2020}.
We classify the contributions of the VAG conductivity by using the similarity to the NLE conductivity,
\begin{align}
  \sigma^{\alpha;\beta\gamma}_{\mathrm{s}} = \sigma^{\alpha;\beta\gamma}_{\mathrm{s,Drude}} + \sigma^{\alpha;\beta\gamma}_{\mathrm{s,BCP}} + \sigma^{\alpha;\beta\gamma}_{\mathrm{s,NRSF}}, 
\end{align}
where $\sigma_{\mathrm{s,Drude}}$, $\sigma_{\mathrm{s,BCP}}$, and $\sigma_{\mathrm{s,NRSF}}$ are the Drude~\cite{Watanabe2020, Watanabe2022}, Berry connection polarizability~\cite{Gao2014, Das2023, Kaplan2024}, and nonreciprocal superfluid density (NRSF) terms~\cite{Watanabe2022}, respectively.
In our model, the Berry curvature dipole term vanishes due to $\mathrm{C_{3}}$ symmetry~\cite{Sodemann2015}.

\emph{Nonreciprocal pseudo-superfluid density.---}
We discuss the NRSF term for VAG conductivity $\sigma^{\alpha;\beta\gamma}_{\mathrm{s,NRSF}}$ focusing on its similarity to that for NLE conductivity $\sigma^{\alpha;\beta\gamma}_{\mathrm{e,NRSF}}$ in superconductors.
The superfluidity and macroscopic coherence of the superconducting state is characterized by the superfluid density $\rho^{\alpha\beta}=\lim_{\bm{\lambda}\rightarrow\bm{0}}\partial_{\lambda_{\alpha}}\partial_{\lambda_{\beta}}F[\bm{A}=\bm{\lambda}]$, where $F_{\bm{\lambda}}$ is the free energy and $\bm{\lambda}$ is the variational parameters corresponding to the electromagnetic vector potential $\bm{A}$.
In NLE responses, the NRSF term $\sigma_{\mathrm{e, NRSF}}$ is associted with the nonreciprocal correction to the superfluid density, which we call the NRSF $f^{\alpha\beta\gamma}=\lim_{\bm{\lambda}\rightarrow\bm{0}}\partial_{\lambda_{\alpha}}\partial_{\lambda_{\beta}}\partial_{\lambda_{\gamma}}F[\bm{A}=\bm{\lambda}]$.
The NRSF term $\sigma_{\mathrm{e, NRSF}}$ of NLE conductivity is written by
\begin{align}
    \sigma^{\alpha;\beta\gamma}_{\mathrm{e,NRSF}} = -\frac{1}{2\Omega^{2}}f^{\alpha\beta\gamma}.
    \label{eq:NLE_NRSF}
\end{align}
Thus, the NRSF term originates from the unique macroscopic coherence in superconductors and is most divergent at low frequencies.
In the VAG conductivity, the NRSF term is given by the nonreciprocal pseudo-superfluid density (NRPSF) $f^{\alpha;\beta\gamma}_{\mathrm{s}}$,
\begin{align}
    \sigma^{\alpha;\beta\gamma}_{\mathrm{s,NRSF}}=-\frac{1}{2\Omega^{2}}f^{\alpha;\beta\gamma}_{\mathrm{s}},
    \label{eq:VAGE_NRSF}
\end{align}
where $f^{\alpha;\beta\gamma}_{\mathrm{s}}$ is defined as
\begin{align}
    f^{\alpha;\beta\gamma}_{\mathrm{s}} = \lim_{\bm{\lambda}, \tilde{\bm{\lambda}}\rightarrow\bm{0}}\partial_{\lambda_{\alpha}}\partial_{\tilde{\lambda}_{\beta}}\partial_{\tilde{\lambda}_{\gamma}}F[\bm{A}_{\mathrm{v}}=\bm{\lambda}, \bm{A}_{\mathrm{s}}=\tilde{\bm{\lambda}}].
\end{align}
The valley gauge field $\bm{A}_{\mathrm{v}}$ is coupled to the Hamiltonian by $\mathcal{H}^{\alpha\beta}_{\xi}(\bm{k})\rightarrow \mathcal{H}^{\alpha\beta}_{\xi}(\bm{k}+\frac{e}{\hbar}\xi \bm{A}_{\mathrm{v}})$ for all the matrix elements $(\alpha, \beta)$, while the coupling constants of the pseudogauge field $\bm{A}_{\mathrm{s}}$ are different between the types of matrix elements [Table~\ref{tab:placeholder}] when $r\neq 1$. 
The similarity between Eqs.~\eqref{eq:NLE_NRSF} and \eqref{eq:VAGE_NRSF} demonstrates that the NRSF term in VAG conductivity represents an anomalous contribution resulting from macroscopic coherence.

The analytical formulas for the NLE and VAG conductivities are obtained using two methods: the density matrix formula and the Green's function method~\cite{Watanabe2022, Watanabe2020, supplement}. 
In the Green's function method, the NRSF and NRPSF are expressed in terms of the Green's functions~\cite{Watanabe2022, supplement}.
When neglecting the vertex correction, the effect of impurity scattering is incorporated into the self-energy. 
Thus, the $\Omega^{-2}$ frequency dependence of the NRSF terms holds even in the presence of the impurity scattering when the NRSF or NRPSF is finite. 
In other words, the power-law divergence at low frequencies is an intrinsic property of quantum condensed states.
In this Letter, numerical calculations are performed based on the density matrix formula for computational simplicity.

\emph{NRPSF and valley gauge symmetry breaking.---}
Next, we discuss the relation between the NRPSF $f^{\alpha;\beta\gamma}_{\mathrm s}$ and the IVC order.
Before a detailed discussion, we briefly review the relation between the NRSF $f^{\alpha\beta\gamma}$ and the superconducting order.
Superconducting states are described by the Bogoliubov-de Genne (BdG) Hamiltonian,
\begin{align}
    H_{\mathrm{BdG}}(\bm{k}, \bm{A})=
    \begin{pmatrix}
        H_{\mathrm{N}}(\bm{k}-\frac{e}{\hbar}\bm{A}) & \Delta_{\mathrm{SC}} \\
        \Delta_{\mathrm{SC}}^{\dagger} & H_{\mathrm{N}}(-\bm{k}-\frac{e}{\hbar}\bm{A})^{\top}
    \end{pmatrix},
\end{align}
where the vector potential $\bm{A}$ is incorporated by the minimal coupling.
In the normal state ($\Delta_{\mathrm{SC}}=0$), the $\bm{A}$-diffentiation is almost the same as the $\bm{k}$-differentiation. 
Since the free energy is given by
\begin{align}
    F[\bm{A}=\bm{\lambda}]=-\frac{1}{\beta}\int_{\mathrm{BZ}}\frac{d\bm{k}}{(2\pi)^{2}}\mathrm{Tr}\left[\log\left(1+e^{-\beta H_{\mathrm{BdG}}(\bm{k}, \bm{\lambda})}\right)\right], \notag
\end{align}
we can rewrite $f^{\alpha\beta\gamma}$ using the equivalence of $\bm{A}$- and $\bm{k}$- differentiations, and obtain
\begin{align}
    f^{\alpha\beta\gamma} = \int_{\mathrm{BZ}}\frac{d\bm{k}}{(2\pi)^{2}}\partial_{k_{\alpha}}(\cdots) =0.
\end{align}
We used the periodicity of the Brillouin zone (BZ) in the last equation.
However, the $\bm{A}$- and $\bm{k}$-differentiations are not equivalent in the superconducting state, and the NRSF $f^{\alpha\beta\gamma}$ can be finite~\cite{Watanabe2022}.

A similar relation exists between the NRPSF $f^{\alpha;\beta\gamma}_{\mathrm{s}}$ and the IVC order. In the normal state ($\Delta=0$), the differentiation with the valley gauge field is equivalent to the $\bm{k}$-differentiation for the mean-field Hamiltonian, Eq.~(\ref{eq:MF_Hamiltonian}). Thus, we can show that  $f^{\alpha;\beta\gamma}_{\mathrm{s}}$ almost vanishes as
\begin{align}
    f^{\alpha;\beta\gamma}_{\mathrm{s}} = \int_{D}\frac{d\bm{k}}{(2\pi)^{2}}\partial_{k_{\alpha}}(\cdots).
    \label{eq:total_differential}
\end{align}
We restrict the integral range to the momentum-space domain $D$ because the valley charge is defined there.
Since the domain $D$ does not satisfy periodicity, the NRPSF $f^{\alpha;\beta\gamma}_{\mathrm{s}}$ does not exactly vanish even in the normal state. 
However, our numerical results later show that the magnitude of $f^{\alpha;\beta\gamma}_{\mathrm{s}}$ in the normal state is tiny and becomes smaller as the domain $D$ spreads.
Furthermore, the valley charge is not well-defined at momenta far from the valley points. Therefore, the valley current due to the NRPSF is probably meaningless in the normal state.
In the IVC state, the IVC order forbids rewriting the integrand into the total differential with ${\bm k}$ as in Eq.~(\ref{eq:total_differential}). Therefore, 
a large and physically meaningful NRPSF $f^{\alpha;\beta\gamma}_{\mathrm{s}}$ appears.
Indeed, we numerically demonstrate that the magnitude of $f^{\alpha;\beta\gamma}_{\mathrm{s}}$ in the IVC state is much larger than that in the normal state. The similarity between NRSF and NRPSF confirms the analogy between superconductivity and IVC order.

\emph{Numerical demonstration.---}
Based on the above analysis, we have predicted that the IVC order makes anomalous contributions to the VAG effect, which are divergent at low frequencies. This phenomenon is a counterpart of the nonlinear electric responses in the superconducting state.
Here, we numerically demonstrate the significant impact of the IVC order on the VAG effect.

\begin{figure}[tbp]
 \includegraphics[width=1.0\linewidth]{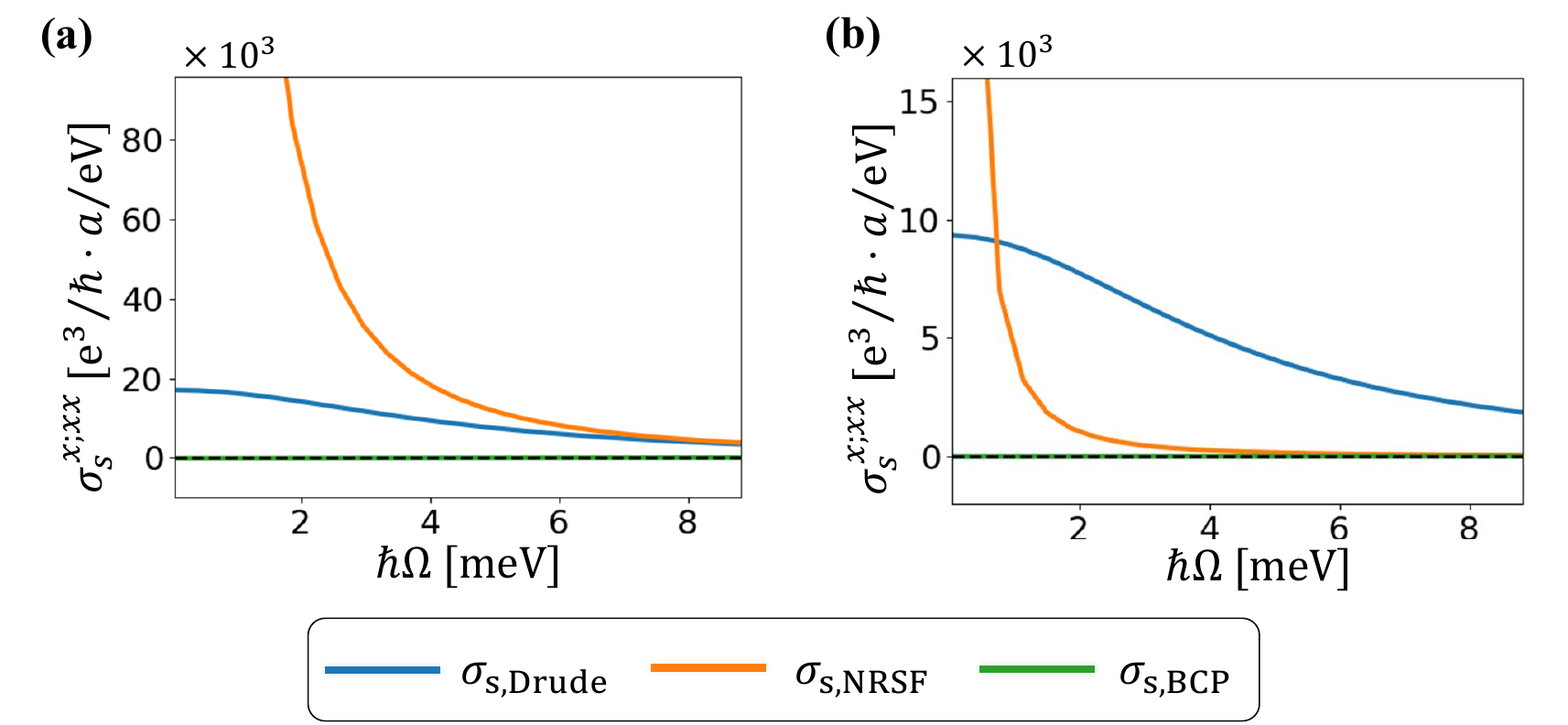}
 \caption{Frequency dependence of the VAG conductivity, which stems from the Drude term $\sigma_{\mathrm{s, Drude}}$, the Berry connection polarizability term $\sigma_{\mathrm{s, BCP}}$, and the NRSF term $\sigma_{\mathrm{s, NRSF}}$. We set (a) $\Delta=30\,\mathrm{meV}$ (IVC state) and (b) $\Delta=0\,\mathrm{meV}$ (normal state).}
 \label{fig:frequency_sigma}
\end{figure}

The VAG conductivity satisfies a relation $\sigma^{x;xx}_{\mathrm{s}}=-\sigma^{x;yy}_{\mathrm{s}}=-\sigma^{y;xy}_{\mathrm{s}}$, and the other components must be zero due to $\mathrm{C_{3v}}$ symmetry.
The frequency dependence of the VAG conductivity $\sigma^{x;xx}_{\mathrm{s}}$ in the IVC state ($\Delta=30\text{ meV}$) is shown in Fig.~\ref{fig:frequency_sigma}(a). Although the typical frequency of the SAW is in $\mu$eV order, we plot the VAG conductivity in the frequency range from zero to meV order, in order to clarify the frequency dependence characteristic of each mechanism of the VAG effects. 
The $\Omega^{-2}$-divergent behavior is unique to the NRSF conductivity, and thus it is the dominant 
VAG effect in the $\mu$eV order frequency region. 

The VAG conductivity in the normal state is plotted in Fig.~\ref{fig:frequency_sigma}(b).
A finite NRSF conductivity appears and becomes dominant in the low-frequency region $\Omega \ll 1\text{ meV}$ due to the $\Omega^{-2}$ divergence. 
However, the magnitude of the NRSF conductivity in the normal state is significantly smaller than that in the IVC state. 
Therefore, the IVC order 
remarkably enhances the low-frequency VAG conductivity.

\begin{figure}[tbp]
 \includegraphics[width=1.0\linewidth]{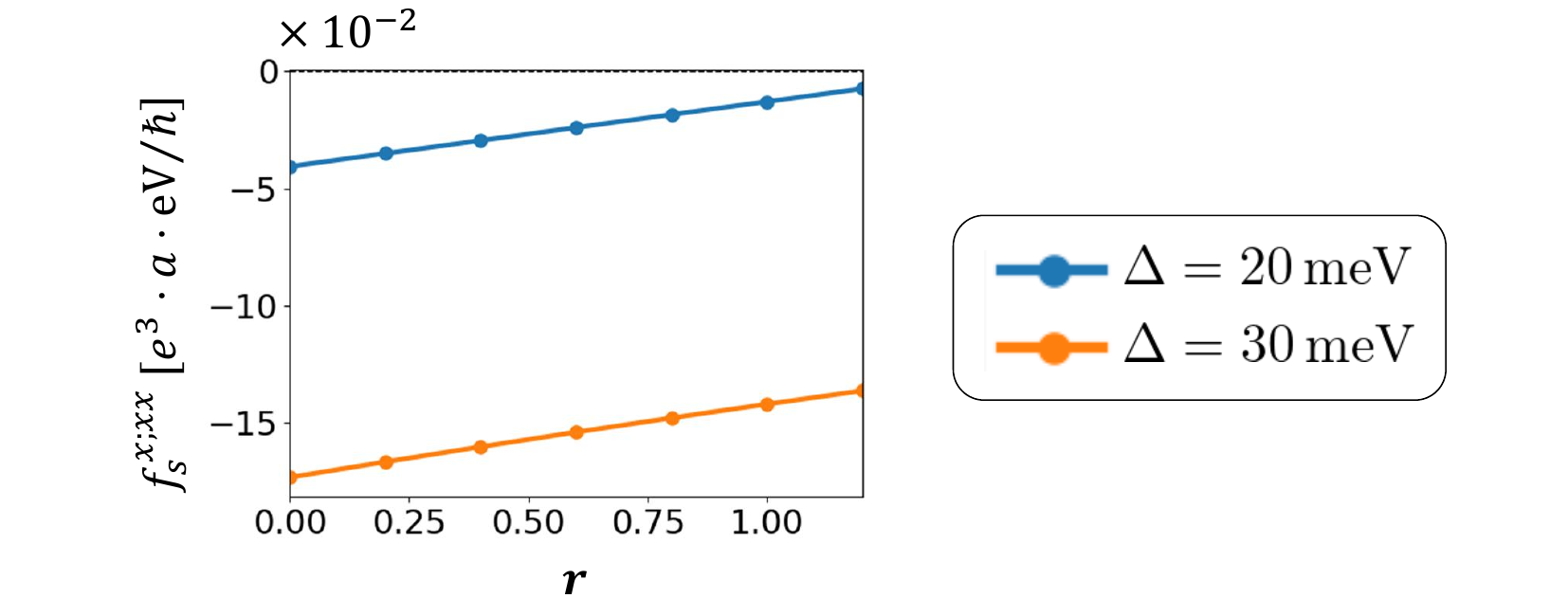}
 \caption{The NRPSF $f^{x;xx}_{\mathrm{s}}$ as a function of $r$ for the IVC order parameter $\Delta=20\,$meV and $30\,$meV.}
 \label{fig:r_chi}
\end{figure}

The NRPSF $f^{\alpha;\beta\gamma}_{\mathrm{s}}$ determines the magnitude of the NRSF term in the VAG conductivity through Eq.~(\ref{eq:VAGE_NRSF}).
Therefore, we here investigate how the NRPSF depends on the parameter $r$, which is the coupling factor for the pseudogauge field $\bm{A}_{\mathrm{s}}$ in type C1 and C2 matrix elements [Table \ref{tab:placeholder}].
Figure~\ref{fig:r_chi} shows the $r$ dependence of the NRPSF $f^{x;xx}_{\mathrm{s}}$ with $\Delta = 20$ and 30 meV.
We obtain finite NRPSF at any $r$. 
Furthermore, the magnitude of $f^{x;xx}_{\mathrm{s}}$ increases with increasing the IVC order parameter $\Delta$.
This implies that the IVC order strongly enhances the VAG conductivity regardless of details of the pseudogauge field.


\begin{figure}[tbp]
 \includegraphics[width=1.0\linewidth]{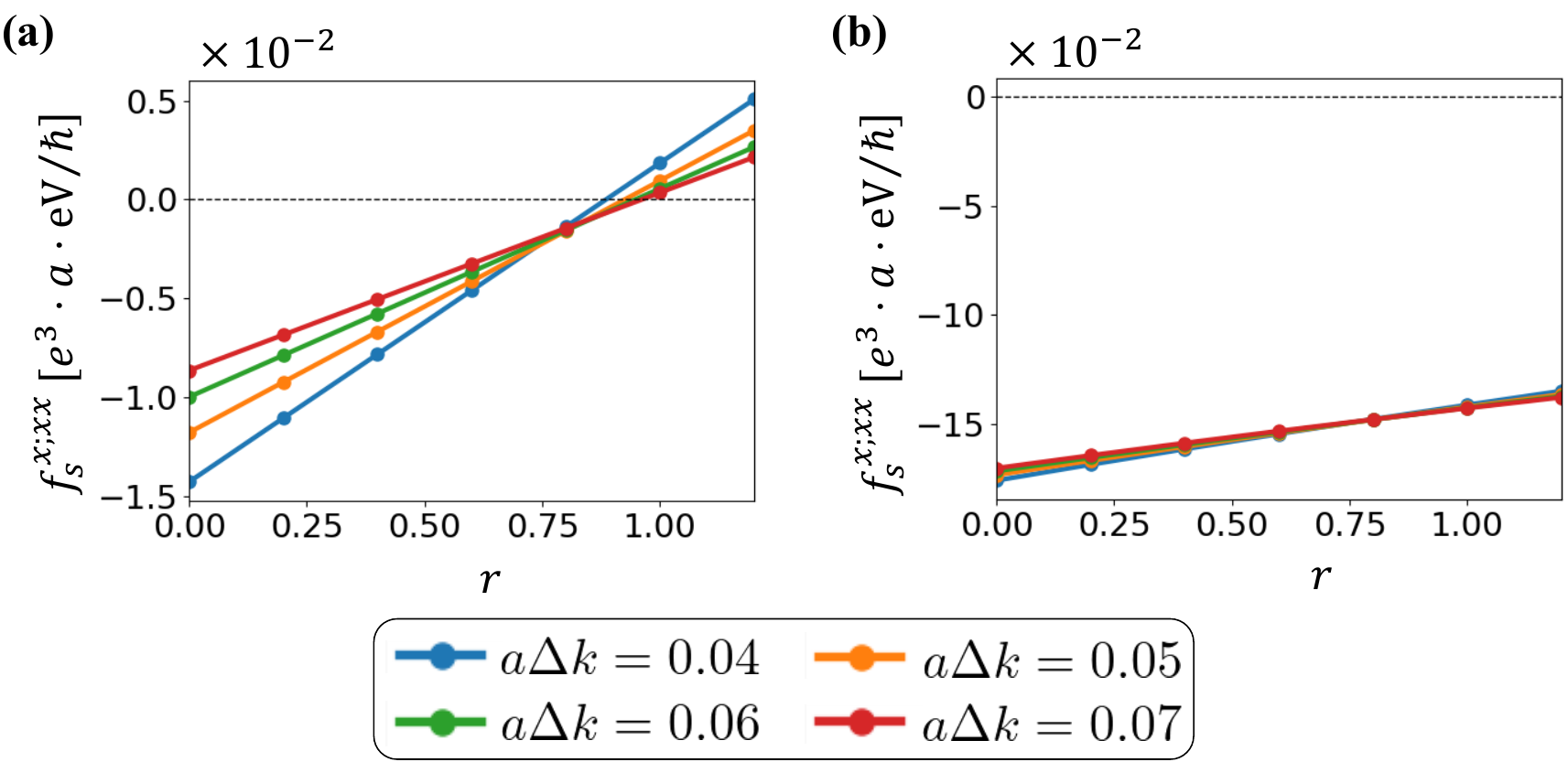}
 \caption{The NRPSF $f^{x;xx}_{s}$ for various $\Delta k$, a parameter specifying the integral domain $D$. We plot the $r$ dependence in (a) the normal state ($\Delta=0\,$meV) and (b) the IVC state ($\Delta=30\,$meV).}
 \label{fig:r_chi_2}
\end{figure}

A finite NRPSF in the normal state, Eq.~(\ref{eq:total_differential}), can be rewritten as the surface integral on the boundary of the domain $D$, whereas this rewriting is forbidden in the IVC state.
Thus, in the normal state the NRPSF is expected to depend heavily on the choice of the domain $D$, while in the IVC state it is free from such ambiguity. 
This expectation is verified by our numerical calculation.
Figure~\ref{fig:r_chi_2} plots the 
NRPSF $f^{\alpha;\beta\gamma}_{\mathrm{s}}$ in the normal and IVC states with varying $\Delta k$, the parameter of the domain size $D$.
In the normal state, the magnitude of 
the NRPSF gets smaller when the domain is enlarged. 
On the other hand, the NRPSF in the IVC state is almost independent of $\Delta k$.
These results support our expectations. Thus, we conclude that the NRPSF is an intrinsic quantity in the IVC state, which verifies the pseudo-superfluidity of the IVC state.
In contrast, the NRPSF in the normal state remains as the boundary term, and its magnitude is tiny. The physical meaning of the boundary term is ambiguous 
because the valley charge is ill-defined far from the valley points.
Thus, the NRPSF in the normal state has no physical significance. 
In any case, the VAG effect can indicate the IVC order via the NRPSF.

\emph{Discussion.---}
Here, we estimate the SAW-driven valley current under a realistic experimental condition. 
For $\Delta\simeq$ 30 meV we obtain $|f^{x;xx}_{\mathrm{s}}|\simeq$ 0.1 $e^{3}\cdot a\cdot \text{eV}/\hbar$. 
The SAWs~\cite{Castro2007, Rummel2021} of the parameters, $u_{L}\simeq$ 2.4 pm $(\simeq0.8u_{z})$, $\beta\simeq$ 3, $a\simeq$ 2.5\AA, $\lambda_{\mathrm{SAW}}=2\pi/Q\simeq$ 10$\mu$m and $\Omega/2\pi\simeq$ 270 MHz produce the pseudo-electric field $|\bm{E}_{\mathrm{s}, 0}|\simeq$ 20 V/m, which gives rise to the valley current of the order of 1 A/m.
When the SAW is generated by the interdigital transducers (IDTs) on the piezoelectric substrates, we can control the frequency of the SAW by varying the pitch of the IDTs~\cite{Nie2023}.
The frequency dependence can be utilized to identify the $\Omega^{-2}$-divergent contribution of the NRSF term.

Nonlocal transport measurements can detect the driven valley currents via the inverse valley Hall effect~\cite{Sui2015, Gorbachev2014, Shimazaki2015, Hung2019, Komatsu2018}.
An alternative and powerful method is Kerr rotation microscopy, one of the optical techniques~\cite{Lee2016, Lee2017}.
Circularly polarized light selectively causes the optical transition in a specific valley due to the selection rules.
Therefore, valley accumulation induces Kerr rotation, and it can be used to probe the valley current.
These transport and optical probes can be applied to atomic layer systems~\cite{Sui2015, Gorbachev2014, Shimazaki2015, Hung2019, Komatsu2018, Lee2016, Lee2017}, platforms of the IVC order.

We comment on valley-charge conservation in the IVC state. 
The IVC ordering breaks the valley gauge symmetry in the mean-field theory.
However, when the interaction Hamiltonian leading to the IVC order is valley-gauge-symmetric, valley-charge conservation holds, as proved analytically.
In realistic graphene materials, the intervalley scattering arises from the electron-phonon coupling~\cite{Banszerus2025} and the short-range impurity potentials~\cite{Peres2010}, but valley currents have been observed unambiguously \cite{Sui2015, Gorbachev2014, Shimazaki2015, Hung2019, Komatsu2018}.
Therefore, the VAG effects in the IVC states can be observed experimentally.


\emph{Conclusion.---}
We have studied SAW-driven valley-current generation in the IVC state. 
We showed that the IVC order significantly enhances valley-current generation via valley-gauge symmetry breaking. 
The $\Omega^{-2}$-divergent contribution to the VAG conductivity is attributed to the NRPSF, which is the third-order derivative of the free energy with respect to the pseudogauge and valley gauge fields.
The NRPSF 
vanishes in the normal state when we assume the valley charge is well-defined in the whole BZ.
Thus, the contribution arising from the NRPSF is unique to the valley-gauge-symmetry-broken IVC state.
The SAW-driven valley-current generation highlights the analogy between the IVC order and superconductivity, and it may be useful for probing the IVC states.
Various classes of two-dimensional materials with honeycomb lattice structure, such as rhombohedral graphene~\cite{Vituri2025, Chatterjee2022, Das2024, Patri2023, Arp2024}, twisted transition metal dichalcogenide~\cite{Munoz2025}, and twisted multilayer graphene~\cite{Po2018,Kwan2021, Nuckolls2023, Lian2021, Zhang2020, Kim2023, Bultinck2020}, are candidates for the IVC state. 
Our results open a promising route to explore exotic response phenomena unique to valley-gauge-symmetry-broken states.

\begin{acknowledgments}
The authors are grateful to T. Matsumoto and Y. Hirobe for fruitful discussions.
This work is supported by JSPS KAKENHI (Grant Numbers JP22H01181, JP22H04933, JP23K17353, JP23K22452, JP24K21530, JP24H00007, JP25H01249).
H.T. is supported by JSPS KAKENHI (Grants Number JP23KJ1344).
\end{acknowledgments}

\bibliography{reference}

\begin{thebibliography}{61}%
\makeatletter
\providecommand \@ifxundefined [1]{%
 \@ifx{#1\undefined}
}%
\providecommand \@ifnum [1]{%
 \ifnum #1\expandafter \@firstoftwo
 \else \expandafter \@secondoftwo
 \fi
}%
\providecommand \@ifx [1]{%
 \ifx #1\expandafter \@firstoftwo
 \else \expandafter \@secondoftwo
 \fi
}%
\providecommand \natexlab [1]{#1}%
\providecommand \enquote  [1]{``#1''}%
\providecommand \bibnamefont  [1]{#1}%
\providecommand \bibfnamefont [1]{#1}%
\providecommand \citenamefont [1]{#1}%
\providecommand \href@noop [0]{\@secondoftwo}%
\providecommand \href [0]{\begingroup \@sanitize@url \@href}%
\providecommand \@href[1]{\@@startlink{#1}\@@href}%
\providecommand \@@href[1]{\endgroup#1\@@endlink}%
\providecommand \@sanitize@url [0]{\catcode `\\12\catcode `\$12\catcode `\&12\catcode `\#12\catcode `\^12\catcode `\_12\catcode `\%12\relax}%
\providecommand \@@startlink[1]{}%
\providecommand \@@endlink[0]{}%
\providecommand \url  [0]{\begingroup\@sanitize@url \@url }%
\providecommand \@url [1]{\endgroup\@href {#1}{\urlprefix }}%
\providecommand \urlprefix  [0]{URL }%
\providecommand \Eprint [0]{\href }%
\providecommand \doibase [0]{https://doi.org/}%
\providecommand \selectlanguage [0]{\@gobble}%
\providecommand \bibinfo  [0]{\@secondoftwo}%
\providecommand \bibfield  [0]{\@secondoftwo}%
\providecommand \translation [1]{[#1]}%
\providecommand \BibitemOpen [0]{}%
\providecommand \bibitemStop [0]{}%
\providecommand \bibitemNoStop [0]{.\EOS\space}%
\providecommand \EOS [0]{\spacefactor3000\relax}%
\providecommand \BibitemShut  [1]{\csname bibitem#1\endcsname}%
\let\auto@bib@innerbib\@empty
\bibitem [{\citenamefont {Liu}\ \emph {et~al.}(2022)\citenamefont {Liu}, \citenamefont {Farahi}, \citenamefont {Chiu}, \citenamefont {Papic}, \citenamefont {Watanabe}, \citenamefont {Taniguchi}, \citenamefont {Zaletel},\ and\ \citenamefont {Yazdani}}]{Liu2022}%
  \BibitemOpen
  \bibfield  {author} {\bibinfo {author} {\bibfnamefont {X.}~\bibnamefont {Liu}}, \bibinfo {author} {\bibfnamefont {G.}~\bibnamefont {Farahi}}, \bibinfo {author} {\bibfnamefont {C.-L.}\ \bibnamefont {Chiu}}, \bibinfo {author} {\bibfnamefont {Z.}~\bibnamefont {Papic}}, \bibinfo {author} {\bibfnamefont {K.}~\bibnamefont {Watanabe}}, \bibinfo {author} {\bibfnamefont {T.}~\bibnamefont {Taniguchi}}, \bibinfo {author} {\bibfnamefont {M.~P.}\ \bibnamefont {Zaletel}},\ and\ \bibinfo {author} {\bibfnamefont {A.}~\bibnamefont {Yazdani}},\ }\bibfield  {title} {\bibinfo {title} {Visualizing broken symmetry and topological defects in a quantum hall ferromagnet},\ }\href {https://doi.org/10.1126/science.abm3770} {\bibfield  {journal} {\bibinfo  {journal} {Science}\ }\textbf {\bibinfo {volume} {375}},\ \bibinfo {pages} {321} (\bibinfo {year} {2022})},\ \Eprint {https://arxiv.org/abs/https://www.science.org/doi/pdf/10.1126/science.abm3770} {https://www.science.org/doi/pdf/10.1126/science.abm3770} \BibitemShut {NoStop}%
\bibitem [{\citenamefont {Farahi}\ \emph {et~al.}(2023)\citenamefont {Farahi}, \citenamefont {Chiu}, \citenamefont {Liu}, \citenamefont {Papic}, \citenamefont {Watanabe}, \citenamefont {Taniguchi}, \citenamefont {Zaletel},\ and\ \citenamefont {Yazdani}}]{Farahi2023}%
  \BibitemOpen
  \bibfield  {author} {\bibinfo {author} {\bibfnamefont {G.}~\bibnamefont {Farahi}}, \bibinfo {author} {\bibfnamefont {C.-L.}\ \bibnamefont {Chiu}}, \bibinfo {author} {\bibfnamefont {X.}~\bibnamefont {Liu}}, \bibinfo {author} {\bibfnamefont {Z.}~\bibnamefont {Papic}}, \bibinfo {author} {\bibfnamefont {K.}~\bibnamefont {Watanabe}}, \bibinfo {author} {\bibfnamefont {T.}~\bibnamefont {Taniguchi}}, \bibinfo {author} {\bibfnamefont {M.~P.}\ \bibnamefont {Zaletel}},\ and\ \bibinfo {author} {\bibfnamefont {A.}~\bibnamefont {Yazdani}},\ }\bibfield  {title} {\bibinfo {title} {Broken symmetries and excitation spectra of interacting electrons in partially filled landau levels},\ }\href {https://doi.org/10.1038/s41567-023-02126-z} {\bibfield  {journal} {\bibinfo  {journal} {Nature Physics}\ }\textbf {\bibinfo {volume} {19}},\ \bibinfo {pages} {1482} (\bibinfo {year} {2023})}\BibitemShut {NoStop}%
\bibitem [{\citenamefont {Fan}\ \emph {et~al.}(2025)\citenamefont {Fan}, \citenamefont {Guo}, \citenamefont {Li}, \citenamefont {Qiu}, \citenamefont {Zhou}, \citenamefont {Zhang}, \citenamefont {Liu}, \citenamefont {Wu},\ and\ \citenamefont {Fu}}]{Fan2025}%
  \BibitemOpen
  \bibfield  {author} {\bibinfo {author} {\bibfnamefont {K.}~\bibnamefont {Fan}}, \bibinfo {author} {\bibfnamefont {T.-F.}\ \bibnamefont {Guo}}, \bibinfo {author} {\bibfnamefont {B.}~\bibnamefont {Li}}, \bibinfo {author} {\bibfnamefont {W.-X.}\ \bibnamefont {Qiu}}, \bibinfo {author} {\bibfnamefont {J.-W.}\ \bibnamefont {Zhou}}, \bibinfo {author} {\bibfnamefont {W.-H.}\ \bibnamefont {Zhang}}, \bibinfo {author} {\bibfnamefont {C.-F.}\ \bibnamefont {Liu}}, \bibinfo {author} {\bibfnamefont {F.}~\bibnamefont {Wu}},\ and\ \bibinfo {author} {\bibfnamefont {Y.-S.}\ \bibnamefont {Fu}},\ }\bibfield  {title} {\bibinfo {title} {Visualization of intervalley coherent phase in ${\mathrm{ptse}}_{2}/\text{bilayer}$ graphene heterojunction},\ }\href {https://doi.org/10.1103/mbjb-x6mn} {\bibfield  {journal} {\bibinfo  {journal} {Phys. Rev. Lett.}\ }\textbf {\bibinfo {volume} {135}},\ \bibinfo {pages} {066201} (\bibinfo {year} {2025})}\BibitemShut {NoStop}%
\bibitem [{\citenamefont {Liao}\ \emph {et~al.}(2025)\citenamefont {Liao}, \citenamefont {Wang}, \citenamefont {Zhang}, \citenamefont {Zhang}, \citenamefont {Tong}, \citenamefont {Zhang}, \citenamefont {Cai}, \citenamefont {Tian}, \citenamefont {Hu}, \citenamefont {Zhang}, \citenamefont {Zhang}, \citenamefont {Qin},\ and\ \citenamefont {Yin}}]{Liao2025}%
  \BibitemOpen
  \bibfield  {author} {\bibinfo {author} {\bibfnamefont {W.-Y.}\ \bibnamefont {Liao}}, \bibinfo {author} {\bibfnamefont {W.-X.}\ \bibnamefont {Wang}}, \bibinfo {author} {\bibfnamefont {S.}~\bibnamefont {Zhang}}, \bibinfo {author} {\bibfnamefont {Y.}~\bibnamefont {Zhang}}, \bibinfo {author} {\bibfnamefont {L.-H.}\ \bibnamefont {Tong}}, \bibinfo {author} {\bibfnamefont {W.}~\bibnamefont {Zhang}}, \bibinfo {author} {\bibfnamefont {H.}~\bibnamefont {Cai}}, \bibinfo {author} {\bibfnamefont {Y.}~\bibnamefont {Tian}}, \bibinfo {author} {\bibfnamefont {Y.}~\bibnamefont {Hu}}, \bibinfo {author} {\bibfnamefont {L.}~\bibnamefont {Zhang}}, \bibinfo {author} {\bibfnamefont {L.}~\bibnamefont {Zhang}}, \bibinfo {author} {\bibfnamefont {Z.}~\bibnamefont {Qin}},\ and\ \bibinfo {author} {\bibfnamefont {L.-J.}\ \bibnamefont {Yin}},\ }\bibfield  {title} {\bibinfo {title} {Intervalley coherent order in rhombohedral tetralayer graphene on ${\mathrm{mos}}_{2}$},\ }\href {https://doi.org/10.1103/s27c-kmfs} {\bibfield  {journal} {\bibinfo  {journal} {Phys. Rev. Lett.}\ }\textbf {\bibinfo {volume} {135}},\ \bibinfo {pages} {046202} (\bibinfo {year} {2025})}\BibitemShut {NoStop}%
\bibitem [{\citenamefont {Liu}\ \emph {et~al.}(2024)\citenamefont {Liu}, \citenamefont {Gupta}, \citenamefont {Choi}, \citenamefont {Vituri}, \citenamefont {Stoyanov}, \citenamefont {Xiao}, \citenamefont {Wang}, \citenamefont {Zhou}, \citenamefont {Barick}, \citenamefont {Taniguchi}, \citenamefont {Watanabe}, \citenamefont {Yan}, \citenamefont {Berg}, \citenamefont {Young}, \citenamefont {Beidenkopf},\ and\ \citenamefont {Avraham}}]{Liu2024}%
  \BibitemOpen
  \bibfield  {author} {\bibinfo {author} {\bibfnamefont {Y.}~\bibnamefont {Liu}}, \bibinfo {author} {\bibfnamefont {A.}~\bibnamefont {Gupta}}, \bibinfo {author} {\bibfnamefont {Y.}~\bibnamefont {Choi}}, \bibinfo {author} {\bibfnamefont {Y.}~\bibnamefont {Vituri}}, \bibinfo {author} {\bibfnamefont {H.}~\bibnamefont {Stoyanov}}, \bibinfo {author} {\bibfnamefont {J.}~\bibnamefont {Xiao}}, \bibinfo {author} {\bibfnamefont {Y.}~\bibnamefont {Wang}}, \bibinfo {author} {\bibfnamefont {H.}~\bibnamefont {Zhou}}, \bibinfo {author} {\bibfnamefont {B.}~\bibnamefont {Barick}}, \bibinfo {author} {\bibfnamefont {T.}~\bibnamefont {Taniguchi}}, \bibinfo {author} {\bibfnamefont {K.}~\bibnamefont {Watanabe}}, \bibinfo {author} {\bibfnamefont {B.}~\bibnamefont {Yan}}, \bibinfo {author} {\bibfnamefont {E.}~\bibnamefont {Berg}}, \bibinfo {author} {\bibfnamefont {A.~F.}\ \bibnamefont {Young}}, \bibinfo {author} {\bibfnamefont {H.}~\bibnamefont {Beidenkopf}},\ and\ \bibinfo {author} {\bibfnamefont {N.}~\bibnamefont {Avraham}},\ }\href {https://arxiv.org/abs/2411.11163} {\bibinfo {title} {Visualizing incommensurate inter-valley coherent states in rhombohedral trilayer graphene}} (\bibinfo {year} {2024}),\ \Eprint {https://arxiv.org/abs/2411.11163} {arXiv:2411.11163 [cond-mat.mes-hall]} \BibitemShut {NoStop}%
\bibitem [{\citenamefont {C\ifmmode \u{a}\else \u{a}\fi{}lug\ifmmode~\u{a}\else \u{a}\fi{}ru}\ \emph {et~al.}(2022)\citenamefont {C\ifmmode \u{a}\else \u{a}\fi{}lug\ifmmode~\u{a}\else \u{a}\fi{}ru}, \citenamefont {Regnault}, \citenamefont {Oh}, \citenamefont {Nuckolls}, \citenamefont {Wong}, \citenamefont {Lee}, \citenamefont {Yazdani}, \citenamefont {Vafek},\ and\ \citenamefont {Bernevig}}]{Dumitru2022}%
  \BibitemOpen
  \bibfield  {author} {\bibinfo {author} {\bibfnamefont {D.}~\bibnamefont {C\ifmmode \u{a}\else \u{a}\fi{}lug\ifmmode~\u{a}\else \u{a}\fi{}ru}}, \bibinfo {author} {\bibfnamefont {N.}~\bibnamefont {Regnault}}, \bibinfo {author} {\bibfnamefont {M.}~\bibnamefont {Oh}}, \bibinfo {author} {\bibfnamefont {K.~P.}\ \bibnamefont {Nuckolls}}, \bibinfo {author} {\bibfnamefont {D.}~\bibnamefont {Wong}}, \bibinfo {author} {\bibfnamefont {R.~L.}\ \bibnamefont {Lee}}, \bibinfo {author} {\bibfnamefont {A.}~\bibnamefont {Yazdani}}, \bibinfo {author} {\bibfnamefont {O.}~\bibnamefont {Vafek}},\ and\ \bibinfo {author} {\bibfnamefont {B.~A.}\ \bibnamefont {Bernevig}},\ }\bibfield  {title} {\bibinfo {title} {Spectroscopy of twisted bilayer graphene correlated insulators},\ }\href {https://doi.org/10.1103/PhysRevLett.129.117602} {\bibfield  {journal} {\bibinfo  {journal} {Phys. Rev. Lett.}\ }\textbf {\bibinfo {volume} {129}},\ \bibinfo {pages} {117602} (\bibinfo {year} {2022})}\BibitemShut {NoStop}%
\bibitem [{\citenamefont {Hong}\ \emph {et~al.}(2022)\citenamefont {Hong}, \citenamefont {Soejima},\ and\ \citenamefont {Zaletel}}]{Hong2022}%
  \BibitemOpen
  \bibfield  {author} {\bibinfo {author} {\bibfnamefont {J.~P.}\ \bibnamefont {Hong}}, \bibinfo {author} {\bibfnamefont {T.}~\bibnamefont {Soejima}},\ and\ \bibinfo {author} {\bibfnamefont {M.~P.}\ \bibnamefont {Zaletel}},\ }\bibfield  {title} {\bibinfo {title} {Detecting symmetry breaking in magic angle graphene using scanning tunneling microscopy},\ }\href {https://doi.org/10.1103/PhysRevLett.129.147001} {\bibfield  {journal} {\bibinfo  {journal} {Phys. Rev. Lett.}\ }\textbf {\bibinfo {volume} {129}},\ \bibinfo {pages} {147001} (\bibinfo {year} {2022})}\BibitemShut {NoStop}%
\bibitem [{\citenamefont {Coissard}\ \emph {et~al.}(2022)\citenamefont {Coissard}, \citenamefont {Wander}, \citenamefont {Vignaud}, \citenamefont {Grushin}, \citenamefont {Repellin}, \citenamefont {Watanabe}, \citenamefont {Taniguchi}, \citenamefont {Gay}, \citenamefont {Winkelmann}, \citenamefont {Courtois}, \citenamefont {Sellier},\ and\ \citenamefont {Sac{\'e}p{\'e}}}]{Coissard2022}%
  \BibitemOpen
  \bibfield  {author} {\bibinfo {author} {\bibfnamefont {A.}~\bibnamefont {Coissard}}, \bibinfo {author} {\bibfnamefont {D.}~\bibnamefont {Wander}}, \bibinfo {author} {\bibfnamefont {H.}~\bibnamefont {Vignaud}}, \bibinfo {author} {\bibfnamefont {A.~G.}\ \bibnamefont {Grushin}}, \bibinfo {author} {\bibfnamefont {C.}~\bibnamefont {Repellin}}, \bibinfo {author} {\bibfnamefont {K.}~\bibnamefont {Watanabe}}, \bibinfo {author} {\bibfnamefont {T.}~\bibnamefont {Taniguchi}}, \bibinfo {author} {\bibfnamefont {F.}~\bibnamefont {Gay}}, \bibinfo {author} {\bibfnamefont {C.~B.}\ \bibnamefont {Winkelmann}}, \bibinfo {author} {\bibfnamefont {H.}~\bibnamefont {Courtois}}, \bibinfo {author} {\bibfnamefont {H.}~\bibnamefont {Sellier}},\ and\ \bibinfo {author} {\bibfnamefont {B.}~\bibnamefont {Sac{\'e}p{\'e}}},\ }\bibfield  {title} {\bibinfo {title} {Imaging tunable quantum hall broken-symmetry orders in graphene},\ }\href {https://doi.org/10.1038/s41586-022-04513-7} {\bibfield  {journal} {\bibinfo  {journal} {Nature}\ }\textbf {\bibinfo {volume} {605}},\ \bibinfo {pages} {51} (\bibinfo {year} {2022})}\BibitemShut {NoStop}%
\bibitem [{\citenamefont {Nuckolls}\ \emph {et~al.}(2023)\citenamefont {Nuckolls}, \citenamefont {Lee}, \citenamefont {Oh}, \citenamefont {Wong}, \citenamefont {Soejima}, \citenamefont {Hong}, \citenamefont {C{\u{a}}lug{\u{a}}ru}, \citenamefont {Herzog-Arbeitman}, \citenamefont {Bernevig}, \citenamefont {Watanabe}, \citenamefont {Taniguchi}, \citenamefont {Regnault}, \citenamefont {Zaletel},\ and\ \citenamefont {Yazdani}}]{Nuckolls2023}%
  \BibitemOpen
  \bibfield  {author} {\bibinfo {author} {\bibfnamefont {K.~P.}\ \bibnamefont {Nuckolls}}, \bibinfo {author} {\bibfnamefont {R.~L.}\ \bibnamefont {Lee}}, \bibinfo {author} {\bibfnamefont {M.}~\bibnamefont {Oh}}, \bibinfo {author} {\bibfnamefont {D.}~\bibnamefont {Wong}}, \bibinfo {author} {\bibfnamefont {T.}~\bibnamefont {Soejima}}, \bibinfo {author} {\bibfnamefont {J.~P.}\ \bibnamefont {Hong}}, \bibinfo {author} {\bibfnamefont {D.}~\bibnamefont {C{\u{a}}lug{\u{a}}ru}}, \bibinfo {author} {\bibfnamefont {J.}~\bibnamefont {Herzog-Arbeitman}}, \bibinfo {author} {\bibfnamefont {B.~A.}\ \bibnamefont {Bernevig}}, \bibinfo {author} {\bibfnamefont {K.}~\bibnamefont {Watanabe}}, \bibinfo {author} {\bibfnamefont {T.}~\bibnamefont {Taniguchi}}, \bibinfo {author} {\bibfnamefont {N.}~\bibnamefont {Regnault}}, \bibinfo {author} {\bibfnamefont {M.~P.}\ \bibnamefont {Zaletel}},\ and\ \bibinfo {author} {\bibfnamefont {A.}~\bibnamefont {Yazdani}},\ }\bibfield  {title} {\bibinfo {title} {Quantum textures of the many-body wavefunctions in magic-angle graphene},\ }\href {https://doi.org/10.1038/s41586-023-06226-x} {\bibfield  {journal} {\bibinfo  {journal} {Nature}\ }\textbf {\bibinfo {volume} {620}},\ \bibinfo {pages} {525} (\bibinfo {year} {2023})}\BibitemShut {NoStop}%
\bibitem [{\citenamefont {Kim}\ \emph {et~al.}(2023)\citenamefont {Kim}, \citenamefont {Choi}, \citenamefont {Lantagne-Hurtubise}, \citenamefont {Lewandowski}, \citenamefont {Thomson}, \citenamefont {Kong}, \citenamefont {Zhou}, \citenamefont {Baum}, \citenamefont {Zhang}, \citenamefont {Holleis}, \citenamefont {Watanabe}, \citenamefont {Taniguchi}, \citenamefont {Young}, \citenamefont {Alicea},\ and\ \citenamefont {Nadj-Perge}}]{Kim2023}%
  \BibitemOpen
  \bibfield  {author} {\bibinfo {author} {\bibfnamefont {H.}~\bibnamefont {Kim}}, \bibinfo {author} {\bibfnamefont {Y.}~\bibnamefont {Choi}}, \bibinfo {author} {\bibfnamefont {{\'E}.}~\bibnamefont {Lantagne-Hurtubise}}, \bibinfo {author} {\bibfnamefont {C.}~\bibnamefont {Lewandowski}}, \bibinfo {author} {\bibfnamefont {A.}~\bibnamefont {Thomson}}, \bibinfo {author} {\bibfnamefont {L.}~\bibnamefont {Kong}}, \bibinfo {author} {\bibfnamefont {H.}~\bibnamefont {Zhou}}, \bibinfo {author} {\bibfnamefont {E.}~\bibnamefont {Baum}}, \bibinfo {author} {\bibfnamefont {Y.}~\bibnamefont {Zhang}}, \bibinfo {author} {\bibfnamefont {L.}~\bibnamefont {Holleis}}, \bibinfo {author} {\bibfnamefont {K.}~\bibnamefont {Watanabe}}, \bibinfo {author} {\bibfnamefont {T.}~\bibnamefont {Taniguchi}}, \bibinfo {author} {\bibfnamefont {A.~F.}\ \bibnamefont {Young}}, \bibinfo {author} {\bibfnamefont {J.}~\bibnamefont {Alicea}},\ and\ \bibinfo {author} {\bibfnamefont {S.}~\bibnamefont {Nadj-Perge}},\ }\bibfield  {title} {\bibinfo {title} {Imaging inter-valley coherent order in magic-angle twisted trilayer graphene},\ }\href {https://doi.org/10.1038/s41586-023-06663-8} {\bibfield  {journal} {\bibinfo  {journal} {Nature}\ }\textbf {\bibinfo {volume} {623}},\ \bibinfo {pages} {942} (\bibinfo {year} {2023})}\BibitemShut {NoStop}%
\bibitem [{\citenamefont {Po}\ \emph {et~al.}(2018)\citenamefont {Po}, \citenamefont {Zou}, \citenamefont {Vishwanath},\ and\ \citenamefont {Senthil}}]{Po2018}%
  \BibitemOpen
  \bibfield  {author} {\bibinfo {author} {\bibfnamefont {H.~C.}\ \bibnamefont {Po}}, \bibinfo {author} {\bibfnamefont {L.}~\bibnamefont {Zou}}, \bibinfo {author} {\bibfnamefont {A.}~\bibnamefont {Vishwanath}},\ and\ \bibinfo {author} {\bibfnamefont {T.}~\bibnamefont {Senthil}},\ }\bibfield  {title} {\bibinfo {title} {Origin of mott insulating behavior and superconductivity in twisted bilayer graphene},\ }\href {https://doi.org/10.1103/PhysRevX.8.031089} {\bibfield  {journal} {\bibinfo  {journal} {Phys. Rev. X}\ }\textbf {\bibinfo {volume} {8}},\ \bibinfo {pages} {031089} (\bibinfo {year} {2018})}\BibitemShut {NoStop}%
\bibitem [{\citenamefont {Kwan}\ \emph {et~al.}(2021)\citenamefont {Kwan}, \citenamefont {Wagner}, \citenamefont {Soejima}, \citenamefont {Zaletel}, \citenamefont {Simon}, \citenamefont {Parameswaran},\ and\ \citenamefont {Bultinck}}]{Kwan2021}%
  \BibitemOpen
  \bibfield  {author} {\bibinfo {author} {\bibfnamefont {Y.~H.}\ \bibnamefont {Kwan}}, \bibinfo {author} {\bibfnamefont {G.}~\bibnamefont {Wagner}}, \bibinfo {author} {\bibfnamefont {T.}~\bibnamefont {Soejima}}, \bibinfo {author} {\bibfnamefont {M.~P.}\ \bibnamefont {Zaletel}}, \bibinfo {author} {\bibfnamefont {S.~H.}\ \bibnamefont {Simon}}, \bibinfo {author} {\bibfnamefont {S.~A.}\ \bibnamefont {Parameswaran}},\ and\ \bibinfo {author} {\bibfnamefont {N.}~\bibnamefont {Bultinck}},\ }\bibfield  {title} {\bibinfo {title} {Kekul\'e spiral order at all nonzero integer fillings in twisted bilayer graphene},\ }\href {https://doi.org/10.1103/PhysRevX.11.041063} {\bibfield  {journal} {\bibinfo  {journal} {Phys. Rev. X}\ }\textbf {\bibinfo {volume} {11}},\ \bibinfo {pages} {041063} (\bibinfo {year} {2021})}\BibitemShut {NoStop}%
\bibitem [{\citenamefont {Bultinck}\ \emph {et~al.}(2020)\citenamefont {Bultinck}, \citenamefont {Khalaf}, \citenamefont {Liu}, \citenamefont {Chatterjee}, \citenamefont {Vishwanath},\ and\ \citenamefont {Zaletel}}]{Bultinck2020}%
  \BibitemOpen
  \bibfield  {author} {\bibinfo {author} {\bibfnamefont {N.}~\bibnamefont {Bultinck}}, \bibinfo {author} {\bibfnamefont {E.}~\bibnamefont {Khalaf}}, \bibinfo {author} {\bibfnamefont {S.}~\bibnamefont {Liu}}, \bibinfo {author} {\bibfnamefont {S.}~\bibnamefont {Chatterjee}}, \bibinfo {author} {\bibfnamefont {A.}~\bibnamefont {Vishwanath}},\ and\ \bibinfo {author} {\bibfnamefont {M.~P.}\ \bibnamefont {Zaletel}},\ }\bibfield  {title} {\bibinfo {title} {Ground state and hidden symmetry of magic-angle graphene at even integer filling},\ }\href {https://doi.org/10.1103/PhysRevX.10.031034} {\bibfield  {journal} {\bibinfo  {journal} {Phys. Rev. X}\ }\textbf {\bibinfo {volume} {10}},\ \bibinfo {pages} {031034} (\bibinfo {year} {2020})}\BibitemShut {NoStop}%
\bibitem [{\citenamefont {Lian}\ \emph {et~al.}(2021)\citenamefont {Lian}, \citenamefont {Song}, \citenamefont {Regnault}, \citenamefont {Efetov}, \citenamefont {Yazdani},\ and\ \citenamefont {Bernevig}}]{Lian2021}%
  \BibitemOpen
  \bibfield  {author} {\bibinfo {author} {\bibfnamefont {B.}~\bibnamefont {Lian}}, \bibinfo {author} {\bibfnamefont {Z.-D.}\ \bibnamefont {Song}}, \bibinfo {author} {\bibfnamefont {N.}~\bibnamefont {Regnault}}, \bibinfo {author} {\bibfnamefont {D.~K.}\ \bibnamefont {Efetov}}, \bibinfo {author} {\bibfnamefont {A.}~\bibnamefont {Yazdani}},\ and\ \bibinfo {author} {\bibfnamefont {B.~A.}\ \bibnamefont {Bernevig}},\ }\bibfield  {title} {\bibinfo {title} {Twisted bilayer graphene. iv. exact insulator ground states and phase diagram},\ }\href {https://doi.org/10.1103/PhysRevB.103.205414} {\bibfield  {journal} {\bibinfo  {journal} {Phys. Rev. B}\ }\textbf {\bibinfo {volume} {103}},\ \bibinfo {pages} {205414} (\bibinfo {year} {2021})}\BibitemShut {NoStop}%
\bibitem [{\citenamefont {Zhang}\ \emph {et~al.}(2020)\citenamefont {Zhang}, \citenamefont {Jiang}, \citenamefont {Wang},\ and\ \citenamefont {Zhang}}]{Zhang2020}%
  \BibitemOpen
  \bibfield  {author} {\bibinfo {author} {\bibfnamefont {Y.}~\bibnamefont {Zhang}}, \bibinfo {author} {\bibfnamefont {K.}~\bibnamefont {Jiang}}, \bibinfo {author} {\bibfnamefont {Z.}~\bibnamefont {Wang}},\ and\ \bibinfo {author} {\bibfnamefont {F.}~\bibnamefont {Zhang}},\ }\bibfield  {title} {\bibinfo {title} {Correlated insulating phases of twisted bilayer graphene at commensurate filling fractions: A hartree-fock study},\ }\href {https://doi.org/10.1103/PhysRevB.102.035136} {\bibfield  {journal} {\bibinfo  {journal} {Phys. Rev. B}\ }\textbf {\bibinfo {volume} {102}},\ \bibinfo {pages} {035136} (\bibinfo {year} {2020})}\BibitemShut {NoStop}%
\bibitem [{\citenamefont {Vituri}\ \emph {et~al.}(2025)\citenamefont {Vituri}, \citenamefont {Xiao}, \citenamefont {Pareek}, \citenamefont {Holder},\ and\ \citenamefont {Berg}}]{Vituri2025}%
  \BibitemOpen
  \bibfield  {author} {\bibinfo {author} {\bibfnamefont {Y.}~\bibnamefont {Vituri}}, \bibinfo {author} {\bibfnamefont {J.}~\bibnamefont {Xiao}}, \bibinfo {author} {\bibfnamefont {K.}~\bibnamefont {Pareek}}, \bibinfo {author} {\bibfnamefont {T.}~\bibnamefont {Holder}},\ and\ \bibinfo {author} {\bibfnamefont {E.}~\bibnamefont {Berg}},\ }\bibfield  {title} {\bibinfo {title} {Incommensurate intervalley coherent states in abc graphene: Collective modes and superconductivity},\ }\href {https://doi.org/10.1103/PhysRevB.111.075103} {\bibfield  {journal} {\bibinfo  {journal} {Phys. Rev. B}\ }\textbf {\bibinfo {volume} {111}},\ \bibinfo {pages} {075103} (\bibinfo {year} {2025})}\BibitemShut {NoStop}%
\bibitem [{\citenamefont {Chatterjee}\ \emph {et~al.}(2022)\citenamefont {Chatterjee}, \citenamefont {Wang}, \citenamefont {Berg},\ and\ \citenamefont {Zaletel}}]{Chatterjee2022}%
  \BibitemOpen
  \bibfield  {author} {\bibinfo {author} {\bibfnamefont {S.}~\bibnamefont {Chatterjee}}, \bibinfo {author} {\bibfnamefont {T.}~\bibnamefont {Wang}}, \bibinfo {author} {\bibfnamefont {E.}~\bibnamefont {Berg}},\ and\ \bibinfo {author} {\bibfnamefont {M.~P.}\ \bibnamefont {Zaletel}},\ }\bibfield  {title} {\bibinfo {title} {Inter-valley coherent order and isospin fluctuation mediated superconductivity in rhombohedral trilayer graphene},\ }\href {https://doi.org/10.1038/s41467-022-33561-w} {\bibfield  {journal} {\bibinfo  {journal} {Nature Communications}\ }\textbf {\bibinfo {volume} {13}},\ \bibinfo {pages} {6013} (\bibinfo {year} {2022})}\BibitemShut {NoStop}%
\bibitem [{\citenamefont {Das}\ and\ \citenamefont {Huang}(2024)}]{Das2024}%
  \BibitemOpen
  \bibfield  {author} {\bibinfo {author} {\bibfnamefont {M.}~\bibnamefont {Das}}\ and\ \bibinfo {author} {\bibfnamefont {C.}~\bibnamefont {Huang}},\ }\bibfield  {title} {\bibinfo {title} {Quarter-metal phases in multilayer graphene: Ising-xy and annular lifshitz transitions},\ }\href {https://doi.org/10.1103/PhysRevB.110.035103} {\bibfield  {journal} {\bibinfo  {journal} {Phys. Rev. B}\ }\textbf {\bibinfo {volume} {110}},\ \bibinfo {pages} {035103} (\bibinfo {year} {2024})}\BibitemShut {NoStop}%
\bibitem [{\citenamefont {Patri}\ and\ \citenamefont {Senthil}(2023)}]{Patri2023}%
  \BibitemOpen
  \bibfield  {author} {\bibinfo {author} {\bibfnamefont {A.~S.}\ \bibnamefont {Patri}}\ and\ \bibinfo {author} {\bibfnamefont {T.}~\bibnamefont {Senthil}},\ }\bibfield  {title} {\bibinfo {title} {Strong correlations in abc-stacked trilayer graphene: Moir\'e is important},\ }\href {https://doi.org/10.1103/PhysRevB.107.165122} {\bibfield  {journal} {\bibinfo  {journal} {Phys. Rev. B}\ }\textbf {\bibinfo {volume} {107}},\ \bibinfo {pages} {165122} (\bibinfo {year} {2023})}\BibitemShut {NoStop}%
\bibitem [{\citenamefont {Arp}\ \emph {et~al.}(2024)\citenamefont {Arp}, \citenamefont {Sheekey}, \citenamefont {Zhou}, \citenamefont {Tschirhart}, \citenamefont {Patterson}, \citenamefont {Yoo}, \citenamefont {Holleis}, \citenamefont {Redekop}, \citenamefont {Babikyan}, \citenamefont {Xie}, \citenamefont {Xiao}, \citenamefont {Vituri}, \citenamefont {Holder}, \citenamefont {Taniguchi}, \citenamefont {Watanabe}, \citenamefont {Huber}, \citenamefont {Berg},\ and\ \citenamefont {Young}}]{Arp2024}%
  \BibitemOpen
  \bibfield  {author} {\bibinfo {author} {\bibfnamefont {T.}~\bibnamefont {Arp}}, \bibinfo {author} {\bibfnamefont {O.}~\bibnamefont {Sheekey}}, \bibinfo {author} {\bibfnamefont {H.}~\bibnamefont {Zhou}}, \bibinfo {author} {\bibfnamefont {C.~L.}\ \bibnamefont {Tschirhart}}, \bibinfo {author} {\bibfnamefont {C.~L.}\ \bibnamefont {Patterson}}, \bibinfo {author} {\bibfnamefont {H.~M.}\ \bibnamefont {Yoo}}, \bibinfo {author} {\bibfnamefont {L.}~\bibnamefont {Holleis}}, \bibinfo {author} {\bibfnamefont {E.}~\bibnamefont {Redekop}}, \bibinfo {author} {\bibfnamefont {G.}~\bibnamefont {Babikyan}}, \bibinfo {author} {\bibfnamefont {T.}~\bibnamefont {Xie}}, \bibinfo {author} {\bibfnamefont {J.}~\bibnamefont {Xiao}}, \bibinfo {author} {\bibfnamefont {Y.}~\bibnamefont {Vituri}}, \bibinfo {author} {\bibfnamefont {T.}~\bibnamefont {Holder}}, \bibinfo {author} {\bibfnamefont {T.}~\bibnamefont {Taniguchi}}, \bibinfo {author} {\bibfnamefont {K.}~\bibnamefont {Watanabe}}, \bibinfo {author} {\bibfnamefont {M.~E.}\ \bibnamefont {Huber}}, \bibinfo {author} {\bibfnamefont {E.}~\bibnamefont {Berg}},\ and\ \bibinfo {author} {\bibfnamefont {A.~F.}\ \bibnamefont {Young}},\ }\bibfield  {title} {\bibinfo {title} {Intervalley coherence and intrinsic spin--orbit coupling in rhombohedral trilayer graphene},\ }\href {https://doi.org/10.1038/s41567-024-02560-7} {\bibfield  {journal} {\bibinfo  {journal} {Nature Physics}\ }\textbf {\bibinfo {volume} {20}},\ \bibinfo {pages} {1413} (\bibinfo {year} {2024})}\BibitemShut {NoStop}%
\bibitem [{\citenamefont {Mu\~noz Segovia}\ \emph {et~al.}(2025)\citenamefont {Mu\~noz Segovia}, \citenamefont {Cr\'epel}, \citenamefont {Queiroz},\ and\ \citenamefont {Millis}}]{Munoz2025}%
  \BibitemOpen
  \bibfield  {author} {\bibinfo {author} {\bibfnamefont {D.}~\bibnamefont {Mu\~noz Segovia}}, \bibinfo {author} {\bibfnamefont {V.}~\bibnamefont {Cr\'epel}}, \bibinfo {author} {\bibfnamefont {R.}~\bibnamefont {Queiroz}},\ and\ \bibinfo {author} {\bibfnamefont {A.~J.}\ \bibnamefont {Millis}},\ }\bibfield  {title} {\bibinfo {title} {Twist-angle evolution of the intervalley-coherent antiferromagnet in twisted ${\mathrm{wse}}_{2}$},\ }\href {https://doi.org/10.1103/m6c2-yd5j} {\bibfield  {journal} {\bibinfo  {journal} {Phys. Rev. B}\ }\textbf {\bibinfo {volume} {112}},\ \bibinfo {pages} {085111} (\bibinfo {year} {2025})}\BibitemShut {NoStop}%
\bibitem [{\citenamefont {Das}\ and\ \citenamefont {Huang}(2025)}]{Das2025}%
  \BibitemOpen
  \bibfield  {author} {\bibinfo {author} {\bibfnamefont {M.}~\bibnamefont {Das}}\ and\ \bibinfo {author} {\bibfnamefont {C.}~\bibnamefont {Huang}},\ }\href {https://arxiv.org/abs/2503.22003} {\bibinfo {title} {Momentum-space ac josephson effect and intervalley coherence in multilayer graphene}} (\bibinfo {year} {2025}),\ \Eprint {https://arxiv.org/abs/2503.22003} {arXiv:2503.22003 [cond-mat.mes-hall]} \BibitemShut {NoStop}%
\bibitem [{\citenamefont {Wei}\ \emph {et~al.}(2025)\citenamefont {Wei}, \citenamefont {Zeng},\ and\ \citenamefont {MacDonald}}]{Wei2025}%
  \BibitemOpen
  \bibfield  {author} {\bibinfo {author} {\bibfnamefont {N.}~\bibnamefont {Wei}}, \bibinfo {author} {\bibfnamefont {Y.}~\bibnamefont {Zeng}},\ and\ \bibinfo {author} {\bibfnamefont {A.~H.}\ \bibnamefont {MacDonald}},\ }\bibfield  {title} {\bibinfo {title} {Weak localization as a probe of intervalley coherence in graphene multilayers},\ }\href {https://doi.org/10.1103/PhysRevB.111.L241103} {\bibfield  {journal} {\bibinfo  {journal} {Phys. Rev. B}\ }\textbf {\bibinfo {volume} {111}},\ \bibinfo {pages} {L241103} (\bibinfo {year} {2025})}\BibitemShut {NoStop}%
\bibitem [{\citenamefont {Thomson}\ \emph {et~al.}(2022)\citenamefont {Thomson}, \citenamefont {Sorensen}, \citenamefont {Nadj-Perge},\ and\ \citenamefont {Alicea}}]{Thomson2022}%
  \BibitemOpen
  \bibfield  {author} {\bibinfo {author} {\bibfnamefont {A.}~\bibnamefont {Thomson}}, \bibinfo {author} {\bibfnamefont {I.~M.}\ \bibnamefont {Sorensen}}, \bibinfo {author} {\bibfnamefont {S.}~\bibnamefont {Nadj-Perge}},\ and\ \bibinfo {author} {\bibfnamefont {J.}~\bibnamefont {Alicea}},\ }\bibfield  {title} {\bibinfo {title} {Gate-defined wires in twisted bilayer graphene: From electrical detection of intervalley coherence to internally engineered majorana modes},\ }\href {https://doi.org/10.1103/PhysRevB.105.L081405} {\bibfield  {journal} {\bibinfo  {journal} {Phys. Rev. B}\ }\textbf {\bibinfo {volume} {105}},\ \bibinfo {pages} {L081405} (\bibinfo {year} {2022})}\BibitemShut {NoStop}%
\bibitem [{\citenamefont {Nie}\ \emph {et~al.}(2023)\citenamefont {Nie}, \citenamefont {Wu}, \citenamefont {Wang}, \citenamefont {Ban}, \citenamefont {Lei}, \citenamefont {Yi}, \citenamefont {Liu},\ and\ \citenamefont {Liu}}]{Nie2023}%
  \BibitemOpen
  \bibfield  {author} {\bibinfo {author} {\bibfnamefont {X.}~\bibnamefont {Nie}}, \bibinfo {author} {\bibfnamefont {X.}~\bibnamefont {Wu}}, \bibinfo {author} {\bibfnamefont {Y.}~\bibnamefont {Wang}}, \bibinfo {author} {\bibfnamefont {S.}~\bibnamefont {Ban}}, \bibinfo {author} {\bibfnamefont {Z.}~\bibnamefont {Lei}}, \bibinfo {author} {\bibfnamefont {J.}~\bibnamefont {Yi}}, \bibinfo {author} {\bibfnamefont {Y.}~\bibnamefont {Liu}},\ and\ \bibinfo {author} {\bibfnamefont {Y.}~\bibnamefont {Liu}},\ }\bibfield  {title} {\bibinfo {title} {Surface acoustic wave induced phenomena in two-dimensional materials},\ }\href {https://doi.org/10.1039/D2NH00458E} {\bibfield  {journal} {\bibinfo  {journal} {Nanoscale Horiz.}\ }\textbf {\bibinfo {volume} {8}},\ \bibinfo {pages} {158} (\bibinfo {year} {2023})}\BibitemShut {NoStop}%
\bibitem [{\citenamefont {von Oppen}\ \emph {et~al.}(2009)\citenamefont {von Oppen}, \citenamefont {Guinea},\ and\ \citenamefont {Mariani}}]{von_Oppen2009}%
  \BibitemOpen
  \bibfield  {author} {\bibinfo {author} {\bibfnamefont {F.}~\bibnamefont {von Oppen}}, \bibinfo {author} {\bibfnamefont {F.}~\bibnamefont {Guinea}},\ and\ \bibinfo {author} {\bibfnamefont {E.}~\bibnamefont {Mariani}},\ }\bibfield  {title} {\bibinfo {title} {Synthetic electric fields and phonon damping in carbon nanotubes and graphene},\ }\href {https://doi.org/10.1103/PhysRevB.80.075420} {\bibfield  {journal} {\bibinfo  {journal} {Phys. Rev. B}\ }\textbf {\bibinfo {volume} {80}},\ \bibinfo {pages} {075420} (\bibinfo {year} {2009})}\BibitemShut {NoStop}%
\bibitem [{\citenamefont {Vaezi}\ \emph {et~al.}(2013)\citenamefont {Vaezi}, \citenamefont {Abedpour}, \citenamefont {Asgari}, \citenamefont {Cortijo},\ and\ \citenamefont {Vozmediano}}]{Vaezi2013}%
  \BibitemOpen
  \bibfield  {author} {\bibinfo {author} {\bibfnamefont {A.}~\bibnamefont {Vaezi}}, \bibinfo {author} {\bibfnamefont {N.}~\bibnamefont {Abedpour}}, \bibinfo {author} {\bibfnamefont {R.}~\bibnamefont {Asgari}}, \bibinfo {author} {\bibfnamefont {A.}~\bibnamefont {Cortijo}},\ and\ \bibinfo {author} {\bibfnamefont {M.~A.~H.}\ \bibnamefont {Vozmediano}},\ }\bibfield  {title} {\bibinfo {title} {Topological electric current from time-dependent elastic deformations in graphene},\ }\href {https://doi.org/10.1103/PhysRevB.88.125406} {\bibfield  {journal} {\bibinfo  {journal} {Phys. Rev. B}\ }\textbf {\bibinfo {volume} {88}},\ \bibinfo {pages} {125406} (\bibinfo {year} {2013})}\BibitemShut {NoStop}%
\bibitem [{\citenamefont {Sela}\ \emph {et~al.}(2020)\citenamefont {Sela}, \citenamefont {Bloch}, \citenamefont {von Oppen},\ and\ \citenamefont {Shalom}}]{Sela2020}%
  \BibitemOpen
  \bibfield  {author} {\bibinfo {author} {\bibfnamefont {E.}~\bibnamefont {Sela}}, \bibinfo {author} {\bibfnamefont {Y.}~\bibnamefont {Bloch}}, \bibinfo {author} {\bibfnamefont {F.}~\bibnamefont {von Oppen}},\ and\ \bibinfo {author} {\bibfnamefont {M.~B.}\ \bibnamefont {Shalom}},\ }\bibfield  {title} {\bibinfo {title} {Quantum hall response to time-dependent strain gradients in graphene},\ }\href {https://doi.org/10.1103/PhysRevLett.124.026602} {\bibfield  {journal} {\bibinfo  {journal} {Phys. Rev. Lett.}\ }\textbf {\bibinfo {volume} {124}},\ \bibinfo {pages} {026602} (\bibinfo {year} {2020})}\BibitemShut {NoStop}%
\bibitem [{\citenamefont {Zhao}\ \emph {et~al.}(2022)\citenamefont {Zhao}, \citenamefont {Sharma}, \citenamefont {Liang}, \citenamefont {Glasenapp}, \citenamefont {Mourokh}, \citenamefont {Kovalev}, \citenamefont {Huber}, \citenamefont {Prada}, \citenamefont {Tiemann},\ and\ \citenamefont {Blick}}]{Zhao2022}%
  \BibitemOpen
  \bibfield  {author} {\bibinfo {author} {\bibfnamefont {P.}~\bibnamefont {Zhao}}, \bibinfo {author} {\bibfnamefont {C.~H.}\ \bibnamefont {Sharma}}, \bibinfo {author} {\bibfnamefont {R.}~\bibnamefont {Liang}}, \bibinfo {author} {\bibfnamefont {C.}~\bibnamefont {Glasenapp}}, \bibinfo {author} {\bibfnamefont {L.}~\bibnamefont {Mourokh}}, \bibinfo {author} {\bibfnamefont {V.~M.}\ \bibnamefont {Kovalev}}, \bibinfo {author} {\bibfnamefont {P.}~\bibnamefont {Huber}}, \bibinfo {author} {\bibfnamefont {M.}~\bibnamefont {Prada}}, \bibinfo {author} {\bibfnamefont {L.}~\bibnamefont {Tiemann}},\ and\ \bibinfo {author} {\bibfnamefont {R.~H.}\ \bibnamefont {Blick}},\ }\bibfield  {title} {\bibinfo {title} {Acoustically induced giant synthetic hall voltages in graphene},\ }\href {https://doi.org/10.1103/PhysRevLett.128.256601} {\bibfield  {journal} {\bibinfo  {journal} {Phys. Rev. Lett.}\ }\textbf {\bibinfo {volume} {128}},\ \bibinfo {pages} {256601} (\bibinfo {year} {2022})}\BibitemShut {NoStop}%
\bibitem [{\citenamefont {Bhalla}\ \emph {et~al.}(2022)\citenamefont {Bhalla}, \citenamefont {Vignale},\ and\ \citenamefont {Rostami}}]{Bhalla2022}%
  \BibitemOpen
  \bibfield  {author} {\bibinfo {author} {\bibfnamefont {P.}~\bibnamefont {Bhalla}}, \bibinfo {author} {\bibfnamefont {G.}~\bibnamefont {Vignale}},\ and\ \bibinfo {author} {\bibfnamefont {H.}~\bibnamefont {Rostami}},\ }\bibfield  {title} {\bibinfo {title} {Pseudogauge field driven acoustoelectric current in two-dimensional hexagonal dirac materials},\ }\href {https://doi.org/10.1103/PhysRevB.105.125407} {\bibfield  {journal} {\bibinfo  {journal} {Phys. Rev. B}\ }\textbf {\bibinfo {volume} {105}},\ \bibinfo {pages} {125407} (\bibinfo {year} {2022})}\BibitemShut {NoStop}%
\bibitem [{\citenamefont {Cazalilla}\ \emph {et~al.}(2014)\citenamefont {Cazalilla}, \citenamefont {Ochoa},\ and\ \citenamefont {Guinea}}]{Cazalilla2014}%
  \BibitemOpen
  \bibfield  {author} {\bibinfo {author} {\bibfnamefont {M.~A.}\ \bibnamefont {Cazalilla}}, \bibinfo {author} {\bibfnamefont {H.}~\bibnamefont {Ochoa}},\ and\ \bibinfo {author} {\bibfnamefont {F.}~\bibnamefont {Guinea}},\ }\bibfield  {title} {\bibinfo {title} {Quantum spin hall effect in two-dimensional crystals of transition-metal dichalcogenides},\ }\href {https://doi.org/10.1103/PhysRevLett.113.077201} {\bibfield  {journal} {\bibinfo  {journal} {Phys. Rev. Lett.}\ }\textbf {\bibinfo {volume} {113}},\ \bibinfo {pages} {077201} (\bibinfo {year} {2014})}\BibitemShut {NoStop}%
\bibitem [{\citenamefont {Sukhachov}\ and\ \citenamefont {Rostami}(2020)}]{Sukhachov2020}%
  \BibitemOpen
  \bibfield  {author} {\bibinfo {author} {\bibfnamefont {P.~O.}\ \bibnamefont {Sukhachov}}\ and\ \bibinfo {author} {\bibfnamefont {H.}~\bibnamefont {Rostami}},\ }\bibfield  {title} {\bibinfo {title} {Acoustogalvanic effect in dirac and weyl semimetals},\ }\href {https://doi.org/10.1103/PhysRevLett.124.126602} {\bibfield  {journal} {\bibinfo  {journal} {Phys. Rev. Lett.}\ }\textbf {\bibinfo {volume} {124}},\ \bibinfo {pages} {126602} (\bibinfo {year} {2020})}\BibitemShut {NoStop}%
\bibitem [{\citenamefont {Matsumoto}\ \emph {et~al.}(2025)\citenamefont {Matsumoto}, \citenamefont {Sano}, \citenamefont {Yanase},\ and\ \citenamefont {Daido}}]{Matsumoto2025}%
  \BibitemOpen
  \bibfield  {author} {\bibinfo {author} {\bibfnamefont {T.}~\bibnamefont {Matsumoto}}, \bibinfo {author} {\bibfnamefont {R.}~\bibnamefont {Sano}}, \bibinfo {author} {\bibfnamefont {Y.}~\bibnamefont {Yanase}},\ and\ \bibinfo {author} {\bibfnamefont {A.}~\bibnamefont {Daido}},\ }\href {https://arxiv.org/abs/2505.21436} {\bibinfo {title} {Superconducting acoustogalvanic effect in twisted transition metal dichalcogenides}} (\bibinfo {year} {2025}),\ \Eprint {https://arxiv.org/abs/2505.21436} {arXiv:2505.21436 [cond-mat.supr-con]} \BibitemShut {NoStop}%
\bibitem [{\citenamefont {Uchoa}\ and\ \citenamefont {Barlas}(2013)}]{Uchoa2013}%
  \BibitemOpen
  \bibfield  {author} {\bibinfo {author} {\bibfnamefont {B.}~\bibnamefont {Uchoa}}\ and\ \bibinfo {author} {\bibfnamefont {Y.}~\bibnamefont {Barlas}},\ }\bibfield  {title} {\bibinfo {title} {Superconducting states in pseudo-landau-levels of strained graphene},\ }\href {https://doi.org/10.1103/PhysRevLett.111.046604} {\bibfield  {journal} {\bibinfo  {journal} {Phys. Rev. Lett.}\ }\textbf {\bibinfo {volume} {111}},\ \bibinfo {pages} {046604} (\bibinfo {year} {2013})}\BibitemShut {NoStop}%
\bibitem [{\citenamefont {Massarelli}\ \emph {et~al.}(2017)\citenamefont {Massarelli}, \citenamefont {Wachtel}, \citenamefont {Wei},\ and\ \citenamefont {Paramekanti}}]{Massarelli2017}%
  \BibitemOpen
  \bibfield  {author} {\bibinfo {author} {\bibfnamefont {G.}~\bibnamefont {Massarelli}}, \bibinfo {author} {\bibfnamefont {G.}~\bibnamefont {Wachtel}}, \bibinfo {author} {\bibfnamefont {J.~Y.~T.}\ \bibnamefont {Wei}},\ and\ \bibinfo {author} {\bibfnamefont {A.}~\bibnamefont {Paramekanti}},\ }\bibfield  {title} {\bibinfo {title} {Pseudo-landau levels of bogoliubov quasiparticles in strained nodal superconductors},\ }\href {https://doi.org/10.1103/PhysRevB.96.224516} {\bibfield  {journal} {\bibinfo  {journal} {Phys. Rev. B}\ }\textbf {\bibinfo {volume} {96}},\ \bibinfo {pages} {224516} (\bibinfo {year} {2017})}\BibitemShut {NoStop}%
\bibitem [{\citenamefont {Nica}\ and\ \citenamefont {Franz}(2018)}]{Nica2018}%
  \BibitemOpen
  \bibfield  {author} {\bibinfo {author} {\bibfnamefont {E.~M.}\ \bibnamefont {Nica}}\ and\ \bibinfo {author} {\bibfnamefont {M.}~\bibnamefont {Franz}},\ }\bibfield  {title} {\bibinfo {title} {Landau levels from neutral bogoliubov particles in two-dimensional nodal superconductors under strain and doping gradients},\ }\href {https://doi.org/10.1103/PhysRevB.97.024520} {\bibfield  {journal} {\bibinfo  {journal} {Phys. Rev. B}\ }\textbf {\bibinfo {volume} {97}},\ \bibinfo {pages} {024520} (\bibinfo {year} {2018})}\BibitemShut {NoStop}%
\bibitem [{\citenamefont {Nayga}\ \emph {et~al.}(2019)\citenamefont {Nayga}, \citenamefont {Rachel},\ and\ \citenamefont {Vojta}}]{Nayga2019}%
  \BibitemOpen
  \bibfield  {author} {\bibinfo {author} {\bibfnamefont {M.~M.}\ \bibnamefont {Nayga}}, \bibinfo {author} {\bibfnamefont {S.}~\bibnamefont {Rachel}},\ and\ \bibinfo {author} {\bibfnamefont {M.}~\bibnamefont {Vojta}},\ }\bibfield  {title} {\bibinfo {title} {Magnon landau levels and emergent supersymmetry in strained antiferromagnets},\ }\href {https://doi.org/10.1103/PhysRevLett.123.207204} {\bibfield  {journal} {\bibinfo  {journal} {Phys. Rev. Lett.}\ }\textbf {\bibinfo {volume} {123}},\ \bibinfo {pages} {207204} (\bibinfo {year} {2019})}\BibitemShut {NoStop}%
\bibitem [{\citenamefont {Sano}\ \emph {et~al.}(2024)\citenamefont {Sano}, \citenamefont {Ominato},\ and\ \citenamefont {Matsuo}}]{Sano2024}%
  \BibitemOpen
  \bibfield  {author} {\bibinfo {author} {\bibfnamefont {R.}~\bibnamefont {Sano}}, \bibinfo {author} {\bibfnamefont {Y.}~\bibnamefont {Ominato}},\ and\ \bibinfo {author} {\bibfnamefont {M.}~\bibnamefont {Matsuo}},\ }\bibfield  {title} {\bibinfo {title} {Acoustomagnonic spin hall effect in honeycomb antiferromagnets},\ }\href {https://doi.org/10.1103/PhysRevLett.132.236302} {\bibfield  {journal} {\bibinfo  {journal} {Phys. Rev. Lett.}\ }\textbf {\bibinfo {volume} {132}},\ \bibinfo {pages} {236302} (\bibinfo {year} {2024})}\BibitemShut {NoStop}%
\bibitem [{\citenamefont {Schrieffer}(2018)}]{schrieffer2018}%
  \BibitemOpen
  \bibfield  {author} {\bibinfo {author} {\bibfnamefont {J.~R.}\ \bibnamefont {Schrieffer}},\ }\href@noop {} {\emph {\bibinfo {title} {Theory of superconductivity}}}\ (\bibinfo  {publisher} {CRC press},\ \bibinfo {year} {2018})\BibitemShut {NoStop}%
\bibitem [{sup()}]{supplement}%
  \BibitemOpen
  \href@noop {} {}\bibinfo {note} {See Supplemental Materials}\BibitemShut {NoStop}%
\bibitem [{\citenamefont {Zhang}\ \emph {et~al.}(2010)\citenamefont {Zhang}, \citenamefont {Sahu}, \citenamefont {Min},\ and\ \citenamefont {MacDonald}}]{Zhang2010}%
  \BibitemOpen
  \bibfield  {author} {\bibinfo {author} {\bibfnamefont {F.}~\bibnamefont {Zhang}}, \bibinfo {author} {\bibfnamefont {B.}~\bibnamefont {Sahu}}, \bibinfo {author} {\bibfnamefont {H.}~\bibnamefont {Min}},\ and\ \bibinfo {author} {\bibfnamefont {A.~H.}\ \bibnamefont {MacDonald}},\ }\bibfield  {title} {\bibinfo {title} {Band structure of abc-stacked graphene trilayers},\ }\href {https://doi.org/10.1103/PhysRevB.82.035409} {\bibfield  {journal} {\bibinfo  {journal} {Phys. Rev. B}\ }\textbf {\bibinfo {volume} {82}},\ \bibinfo {pages} {035409} (\bibinfo {year} {2010})}\BibitemShut {NoStop}%
\bibitem [{\citenamefont {Jung}\ and\ \citenamefont {MacDonald}(2013)}]{Jung2013}%
  \BibitemOpen
  \bibfield  {author} {\bibinfo {author} {\bibfnamefont {J.}~\bibnamefont {Jung}}\ and\ \bibinfo {author} {\bibfnamefont {A.~H.}\ \bibnamefont {MacDonald}},\ }\bibfield  {title} {\bibinfo {title} {Gapped broken symmetry states in abc-stacked trilayer graphene},\ }\href {https://doi.org/10.1103/PhysRevB.88.075408} {\bibfield  {journal} {\bibinfo  {journal} {Phys. Rev. B}\ }\textbf {\bibinfo {volume} {88}},\ \bibinfo {pages} {075408} (\bibinfo {year} {2013})}\BibitemShut {NoStop}%
\bibitem [{\citenamefont {Zhou}\ \emph {et~al.}(2021)\citenamefont {Zhou}, \citenamefont {Xie}, \citenamefont {Ghazaryan}, \citenamefont {Holder}, \citenamefont {Ehrets}, \citenamefont {Spanton}, \citenamefont {Taniguchi}, \citenamefont {Watanabe}, \citenamefont {Berg}, \citenamefont {Serbyn},\ and\ \citenamefont {Young}}]{Zhou2021}%
  \BibitemOpen
  \bibfield  {author} {\bibinfo {author} {\bibfnamefont {H.}~\bibnamefont {Zhou}}, \bibinfo {author} {\bibfnamefont {T.}~\bibnamefont {Xie}}, \bibinfo {author} {\bibfnamefont {A.}~\bibnamefont {Ghazaryan}}, \bibinfo {author} {\bibfnamefont {T.}~\bibnamefont {Holder}}, \bibinfo {author} {\bibfnamefont {J.~R.}\ \bibnamefont {Ehrets}}, \bibinfo {author} {\bibfnamefont {E.~M.}\ \bibnamefont {Spanton}}, \bibinfo {author} {\bibfnamefont {T.}~\bibnamefont {Taniguchi}}, \bibinfo {author} {\bibfnamefont {K.}~\bibnamefont {Watanabe}}, \bibinfo {author} {\bibfnamefont {E.}~\bibnamefont {Berg}}, \bibinfo {author} {\bibfnamefont {M.}~\bibnamefont {Serbyn}},\ and\ \bibinfo {author} {\bibfnamefont {A.~F.}\ \bibnamefont {Young}},\ }\bibfield  {title} {\bibinfo {title} {Half- and quarter-metals in rhombohedral trilayer graphene},\ }\href {https://doi.org/10.1038/s41586-021-03938-w} {\bibfield  {journal} {\bibinfo  {journal} {Nature}\ }\textbf {\bibinfo {volume} {598}},\ \bibinfo {pages} {429} (\bibinfo {year} {2021})}\BibitemShut {NoStop}%
\bibitem [{\citenamefont {Mandal}\ and\ \citenamefont {Banerjee}(2022)}]{Mandal2022}%
  \BibitemOpen
  \bibfield  {author} {\bibinfo {author} {\bibfnamefont {D.}~\bibnamefont {Mandal}}\ and\ \bibinfo {author} {\bibfnamefont {S.}~\bibnamefont {Banerjee}},\ }\bibfield  {title} {\bibinfo {title} {Surface acoustic wave (saw) sensors: Physics, materials, and applications},\ }\bibfield  {journal} {\bibinfo  {journal} {Sensors}\ }\textbf {\bibinfo {volume} {22}},\ \href {https://doi.org/10.3390/s22030820} {10.3390/s22030820} (\bibinfo {year} {2022})\BibitemShut {NoStop}%
\bibitem [{\citenamefont {Lee}\ \emph {et~al.}(2016)\citenamefont {Lee}, \citenamefont {Mak},\ and\ \citenamefont {Shan}}]{Lee2016}%
  \BibitemOpen
  \bibfield  {author} {\bibinfo {author} {\bibfnamefont {J.}~\bibnamefont {Lee}}, \bibinfo {author} {\bibfnamefont {K.~F.}\ \bibnamefont {Mak}},\ and\ \bibinfo {author} {\bibfnamefont {J.}~\bibnamefont {Shan}},\ }\bibfield  {title} {\bibinfo {title} {Electrical control of the valley hall effect in bilayer mos2 transistors},\ }\href {https://doi.org/10.1038/nnano.2015.337} {\bibfield  {journal} {\bibinfo  {journal} {Nature Nanotechnology}\ }\textbf {\bibinfo {volume} {11}},\ \bibinfo {pages} {421} (\bibinfo {year} {2016})}\BibitemShut {NoStop}%
\bibitem [{\citenamefont {Lee}\ \emph {et~al.}(2017)\citenamefont {Lee}, \citenamefont {Wang}, \citenamefont {Xie}, \citenamefont {Mak},\ and\ \citenamefont {Shan}}]{Lee2017}%
  \BibitemOpen
  \bibfield  {author} {\bibinfo {author} {\bibfnamefont {J.}~\bibnamefont {Lee}}, \bibinfo {author} {\bibfnamefont {Z.}~\bibnamefont {Wang}}, \bibinfo {author} {\bibfnamefont {H.}~\bibnamefont {Xie}}, \bibinfo {author} {\bibfnamefont {K.~F.}\ \bibnamefont {Mak}},\ and\ \bibinfo {author} {\bibfnamefont {J.}~\bibnamefont {Shan}},\ }\bibfield  {title} {\bibinfo {title} {Valley magnetoelectricity in single-layer mos2},\ }\href {https://doi.org/10.1038/nmat4931} {\bibfield  {journal} {\bibinfo  {journal} {Nature Materials}\ }\textbf {\bibinfo {volume} {16}},\ \bibinfo {pages} {887} (\bibinfo {year} {2017})}\BibitemShut {NoStop}%
\bibitem [{\citenamefont {Watanabe}\ \emph {et~al.}(2022)\citenamefont {Watanabe}, \citenamefont {Daido},\ and\ \citenamefont {Yanase}}]{Watanabe2022}%
  \BibitemOpen
  \bibfield  {author} {\bibinfo {author} {\bibfnamefont {H.}~\bibnamefont {Watanabe}}, \bibinfo {author} {\bibfnamefont {A.}~\bibnamefont {Daido}},\ and\ \bibinfo {author} {\bibfnamefont {Y.}~\bibnamefont {Yanase}},\ }\bibfield  {title} {\bibinfo {title} {Nonreciprocal optical response in parity-breaking superconductors},\ }\href {https://doi.org/10.1103/PhysRevB.105.024308} {\bibfield  {journal} {\bibinfo  {journal} {Phys. Rev. B}\ }\textbf {\bibinfo {volume} {105}},\ \bibinfo {pages} {024308} (\bibinfo {year} {2022})}\BibitemShut {NoStop}%
\bibitem [{\citenamefont {Watanabe}\ and\ \citenamefont {Yanase}(2020)}]{Watanabe2020}%
  \BibitemOpen
  \bibfield  {author} {\bibinfo {author} {\bibfnamefont {H.}~\bibnamefont {Watanabe}}\ and\ \bibinfo {author} {\bibfnamefont {Y.}~\bibnamefont {Yanase}},\ }\bibfield  {title} {\bibinfo {title} {Nonlinear electric transport in odd-parity magnetic multipole systems: Application to mn-based compounds},\ }\href {https://doi.org/10.1103/PhysRevResearch.2.043081} {\bibfield  {journal} {\bibinfo  {journal} {Phys. Rev. Res.}\ }\textbf {\bibinfo {volume} {2}},\ \bibinfo {pages} {043081} (\bibinfo {year} {2020})}\BibitemShut {NoStop}%
\bibitem [{\citenamefont {Gao}\ \emph {et~al.}(2014)\citenamefont {Gao}, \citenamefont {Yang},\ and\ \citenamefont {Niu}}]{Gao2014}%
  \BibitemOpen
  \bibfield  {author} {\bibinfo {author} {\bibfnamefont {Y.}~\bibnamefont {Gao}}, \bibinfo {author} {\bibfnamefont {S.~A.}\ \bibnamefont {Yang}},\ and\ \bibinfo {author} {\bibfnamefont {Q.}~\bibnamefont {Niu}},\ }\bibfield  {title} {\bibinfo {title} {Field induced positional shift of bloch electrons and its dynamical implications},\ }\href {https://doi.org/10.1103/PhysRevLett.112.166601} {\bibfield  {journal} {\bibinfo  {journal} {Phys. Rev. Lett.}\ }\textbf {\bibinfo {volume} {112}},\ \bibinfo {pages} {166601} (\bibinfo {year} {2014})}\BibitemShut {NoStop}%
\bibitem [{\citenamefont {Das}\ \emph {et~al.}(2023)\citenamefont {Das}, \citenamefont {Lahiri}, \citenamefont {Atencia}, \citenamefont {Culcer},\ and\ \citenamefont {Agarwal}}]{Das2023}%
  \BibitemOpen
  \bibfield  {author} {\bibinfo {author} {\bibfnamefont {K.}~\bibnamefont {Das}}, \bibinfo {author} {\bibfnamefont {S.}~\bibnamefont {Lahiri}}, \bibinfo {author} {\bibfnamefont {R.~B.}\ \bibnamefont {Atencia}}, \bibinfo {author} {\bibfnamefont {D.}~\bibnamefont {Culcer}},\ and\ \bibinfo {author} {\bibfnamefont {A.}~\bibnamefont {Agarwal}},\ }\bibfield  {title} {\bibinfo {title} {Intrinsic nonlinear conductivities induced by the quantum metric},\ }\href {https://doi.org/10.1103/PhysRevB.108.L201405} {\bibfield  {journal} {\bibinfo  {journal} {Phys. Rev. B}\ }\textbf {\bibinfo {volume} {108}},\ \bibinfo {pages} {L201405} (\bibinfo {year} {2023})}\BibitemShut {NoStop}%
\bibitem [{\citenamefont {Kaplan}\ \emph {et~al.}(2024)\citenamefont {Kaplan}, \citenamefont {Holder},\ and\ \citenamefont {Yan}}]{Kaplan2024}%
  \BibitemOpen
  \bibfield  {author} {\bibinfo {author} {\bibfnamefont {D.}~\bibnamefont {Kaplan}}, \bibinfo {author} {\bibfnamefont {T.}~\bibnamefont {Holder}},\ and\ \bibinfo {author} {\bibfnamefont {B.}~\bibnamefont {Yan}},\ }\bibfield  {title} {\bibinfo {title} {Unification of nonlinear anomalous hall effect and nonreciprocal magnetoresistance in metals by the quantum geometry},\ }\href {https://doi.org/10.1103/PhysRevLett.132.026301} {\bibfield  {journal} {\bibinfo  {journal} {Phys. Rev. Lett.}\ }\textbf {\bibinfo {volume} {132}},\ \bibinfo {pages} {026301} (\bibinfo {year} {2024})}\BibitemShut {NoStop}%
\bibitem [{\citenamefont {Sodemann}\ and\ \citenamefont {Fu}(2015)}]{Sodemann2015}%
  \BibitemOpen
  \bibfield  {author} {\bibinfo {author} {\bibfnamefont {I.}~\bibnamefont {Sodemann}}\ and\ \bibinfo {author} {\bibfnamefont {L.}~\bibnamefont {Fu}},\ }\bibfield  {title} {\bibinfo {title} {Quantum nonlinear hall effect induced by berry curvature dipole in time-reversal invariant materials},\ }\href {https://doi.org/10.1103/PhysRevLett.115.216806} {\bibfield  {journal} {\bibinfo  {journal} {Phys. Rev. Lett.}\ }\textbf {\bibinfo {volume} {115}},\ \bibinfo {pages} {216806} (\bibinfo {year} {2015})}\BibitemShut {NoStop}%
\bibitem [{\citenamefont {Castro~Neto}\ and\ \citenamefont {Guinea}(2007)}]{Castro2007}%
  \BibitemOpen
  \bibfield  {author} {\bibinfo {author} {\bibfnamefont {A.~H.}\ \bibnamefont {Castro~Neto}}\ and\ \bibinfo {author} {\bibfnamefont {F.}~\bibnamefont {Guinea}},\ }\bibfield  {title} {\bibinfo {title} {Electron-phonon coupling and raman spectroscopy in graphene},\ }\href {https://doi.org/10.1103/PhysRevB.75.045404} {\bibfield  {journal} {\bibinfo  {journal} {Phys. Rev. B}\ }\textbf {\bibinfo {volume} {75}},\ \bibinfo {pages} {045404} (\bibinfo {year} {2007})}\BibitemShut {NoStop}%
\bibitem [{\citenamefont {Rummel}\ \emph {et~al.}(2021)\citenamefont {Rummel}, \citenamefont {Miroshnik}, \citenamefont {Patriotis}, \citenamefont {Li}, \citenamefont {Sinno}, \citenamefont {Henry}, \citenamefont {Balakrishnan},\ and\ \citenamefont {Han}}]{Rummel2021}%
  \BibitemOpen
  \bibfield  {author} {\bibinfo {author} {\bibfnamefont {B.~D.}\ \bibnamefont {Rummel}}, \bibinfo {author} {\bibfnamefont {L.}~\bibnamefont {Miroshnik}}, \bibinfo {author} {\bibfnamefont {M.}~\bibnamefont {Patriotis}}, \bibinfo {author} {\bibfnamefont {A.}~\bibnamefont {Li}}, \bibinfo {author} {\bibfnamefont {T.~R.}\ \bibnamefont {Sinno}}, \bibinfo {author} {\bibfnamefont {M.~D.}\ \bibnamefont {Henry}}, \bibinfo {author} {\bibfnamefont {G.}~\bibnamefont {Balakrishnan}},\ and\ \bibinfo {author} {\bibfnamefont {S.~M.}\ \bibnamefont {Han}},\ }\bibfield  {title} {\bibinfo {title} {Imaging of surface acoustic waves on gaas using 2d confocal raman microscopy and atomic force microscopy},\ }\href {https://doi.org/10.1063/5.0034572} {\bibfield  {journal} {\bibinfo  {journal} {Applied Physics Letters}\ }\textbf {\bibinfo {volume} {118}},\ \bibinfo {pages} {031602} (\bibinfo {year} {2021})}\BibitemShut {NoStop}%
\bibitem [{\citenamefont {Sui}\ \emph {et~al.}(2015)\citenamefont {Sui}, \citenamefont {Chen}, \citenamefont {Ma}, \citenamefont {Shan}, \citenamefont {Tian}, \citenamefont {Watanabe}, \citenamefont {Taniguchi}, \citenamefont {Jin}, \citenamefont {Yao}, \citenamefont {Xiao},\ and\ \citenamefont {Zhang}}]{Sui2015}%
  \BibitemOpen
  \bibfield  {author} {\bibinfo {author} {\bibfnamefont {M.}~\bibnamefont {Sui}}, \bibinfo {author} {\bibfnamefont {G.}~\bibnamefont {Chen}}, \bibinfo {author} {\bibfnamefont {L.}~\bibnamefont {Ma}}, \bibinfo {author} {\bibfnamefont {W.-Y.}\ \bibnamefont {Shan}}, \bibinfo {author} {\bibfnamefont {D.}~\bibnamefont {Tian}}, \bibinfo {author} {\bibfnamefont {K.}~\bibnamefont {Watanabe}}, \bibinfo {author} {\bibfnamefont {T.}~\bibnamefont {Taniguchi}}, \bibinfo {author} {\bibfnamefont {X.}~\bibnamefont {Jin}}, \bibinfo {author} {\bibfnamefont {W.}~\bibnamefont {Yao}}, \bibinfo {author} {\bibfnamefont {D.}~\bibnamefont {Xiao}},\ and\ \bibinfo {author} {\bibfnamefont {Y.}~\bibnamefont {Zhang}},\ }\bibfield  {title} {\bibinfo {title} {Gate-tunable topological valley transport in bilayer graphene},\ }\href {https://doi.org/10.1038/nphys3485} {\bibfield  {journal} {\bibinfo  {journal} {Nature Physics}\ }\textbf {\bibinfo {volume} {11}},\ \bibinfo {pages} {1027} (\bibinfo {year} {2015})}\BibitemShut {NoStop}%
\bibitem [{\citenamefont {Gorbachev}\ \emph {et~al.}(2014)\citenamefont {Gorbachev}, \citenamefont {Song}, \citenamefont {Yu}, \citenamefont {Kretinin}, \citenamefont {Withers}, \citenamefont {Cao}, \citenamefont {Mishchenko}, \citenamefont {Grigorieva}, \citenamefont {Novoselov}, \citenamefont {Levitov},\ and\ \citenamefont {Geim}}]{Gorbachev2014}%
  \BibitemOpen
  \bibfield  {author} {\bibinfo {author} {\bibfnamefont {R.~V.}\ \bibnamefont {Gorbachev}}, \bibinfo {author} {\bibfnamefont {J.~C.~W.}\ \bibnamefont {Song}}, \bibinfo {author} {\bibfnamefont {G.~L.}\ \bibnamefont {Yu}}, \bibinfo {author} {\bibfnamefont {A.~V.}\ \bibnamefont {Kretinin}}, \bibinfo {author} {\bibfnamefont {F.}~\bibnamefont {Withers}}, \bibinfo {author} {\bibfnamefont {Y.}~\bibnamefont {Cao}}, \bibinfo {author} {\bibfnamefont {A.}~\bibnamefont {Mishchenko}}, \bibinfo {author} {\bibfnamefont {I.~V.}\ \bibnamefont {Grigorieva}}, \bibinfo {author} {\bibfnamefont {K.~S.}\ \bibnamefont {Novoselov}}, \bibinfo {author} {\bibfnamefont {L.~S.}\ \bibnamefont {Levitov}},\ and\ \bibinfo {author} {\bibfnamefont {A.~K.}\ \bibnamefont {Geim}},\ }\bibfield  {title} {\bibinfo {title} {Detecting topological currents in graphene superlattices},\ }\href {https://doi.org/10.1126/science.1254966} {\bibfield  {journal} {\bibinfo  {journal} {Science}\ }\textbf {\bibinfo {volume} {346}},\ \bibinfo {pages} {448} (\bibinfo {year} {2014})},\ \Eprint {https://arxiv.org/abs/https://www.science.org/doi/pdf/10.1126/science.1254966} {https://www.science.org/doi/pdf/10.1126/science.1254966} \BibitemShut {NoStop}%
\bibitem [{\citenamefont {Shimazaki}\ \emph {et~al.}(2015)\citenamefont {Shimazaki}, \citenamefont {Yamamoto}, \citenamefont {Borzenets}, \citenamefont {Watanabe}, \citenamefont {Taniguchi},\ and\ \citenamefont {Tarucha}}]{Shimazaki2015}%
  \BibitemOpen
  \bibfield  {author} {\bibinfo {author} {\bibfnamefont {Y.}~\bibnamefont {Shimazaki}}, \bibinfo {author} {\bibfnamefont {M.}~\bibnamefont {Yamamoto}}, \bibinfo {author} {\bibfnamefont {I.~V.}\ \bibnamefont {Borzenets}}, \bibinfo {author} {\bibfnamefont {K.}~\bibnamefont {Watanabe}}, \bibinfo {author} {\bibfnamefont {T.}~\bibnamefont {Taniguchi}},\ and\ \bibinfo {author} {\bibfnamefont {S.}~\bibnamefont {Tarucha}},\ }\bibfield  {title} {\bibinfo {title} {Generation and detection of pure valley current by electrically induced berry curvature in bilayer graphene},\ }\href {https://doi.org/10.1038/nphys3551} {\bibfield  {journal} {\bibinfo  {journal} {Nature Physics}\ }\textbf {\bibinfo {volume} {11}},\ \bibinfo {pages} {1032} (\bibinfo {year} {2015})}\BibitemShut {NoStop}%
\bibitem [{\citenamefont {Hung}\ \emph {et~al.}(2019)\citenamefont {Hung}, \citenamefont {Camsari}, \citenamefont {Zhang}, \citenamefont {Upadhyaya},\ and\ \citenamefont {Chen}}]{Hung2019}%
  \BibitemOpen
  \bibfield  {author} {\bibinfo {author} {\bibfnamefont {T.~Y.~T.}\ \bibnamefont {Hung}}, \bibinfo {author} {\bibfnamefont {K.~Y.}\ \bibnamefont {Camsari}}, \bibinfo {author} {\bibfnamefont {S.}~\bibnamefont {Zhang}}, \bibinfo {author} {\bibfnamefont {P.}~\bibnamefont {Upadhyaya}},\ and\ \bibinfo {author} {\bibfnamefont {Z.}~\bibnamefont {Chen}},\ }\bibfield  {title} {\bibinfo {title} {Direct observation of valley-coupled topological current in mos<sub>2</sub>},\ }\href {https://doi.org/10.1126/sciadv.aau6478} {\bibfield  {journal} {\bibinfo  {journal} {Science Advances}\ }\textbf {\bibinfo {volume} {5}},\ \bibinfo {pages} {eaau6478} (\bibinfo {year} {2019})},\ \Eprint {https://arxiv.org/abs/https://www.science.org/doi/pdf/10.1126/sciadv.aau6478} {https://www.science.org/doi/pdf/10.1126/sciadv.aau6478} \BibitemShut {NoStop}%
\bibitem [{\citenamefont {Komatsu}\ \emph {et~al.}(2018)\citenamefont {Komatsu}, \citenamefont {Morita}, \citenamefont {Watanabe}, \citenamefont {Tsuya}, \citenamefont {Watanabe}, \citenamefont {Taniguchi},\ and\ \citenamefont {Moriyama}}]{Komatsu2018}%
  \BibitemOpen
  \bibfield  {author} {\bibinfo {author} {\bibfnamefont {K.}~\bibnamefont {Komatsu}}, \bibinfo {author} {\bibfnamefont {Y.}~\bibnamefont {Morita}}, \bibinfo {author} {\bibfnamefont {E.}~\bibnamefont {Watanabe}}, \bibinfo {author} {\bibfnamefont {D.}~\bibnamefont {Tsuya}}, \bibinfo {author} {\bibfnamefont {K.}~\bibnamefont {Watanabe}}, \bibinfo {author} {\bibfnamefont {T.}~\bibnamefont {Taniguchi}},\ and\ \bibinfo {author} {\bibfnamefont {S.}~\bibnamefont {Moriyama}},\ }\bibfield  {title} {\bibinfo {title} {Observation of the quantum valley hall state in ballistic graphene superlattices},\ }\href {https://doi.org/10.1126/sciadv.aaq0194} {\bibfield  {journal} {\bibinfo  {journal} {Science Advances}\ }\textbf {\bibinfo {volume} {4}},\ \bibinfo {pages} {eaaq0194} (\bibinfo {year} {2018})},\ \Eprint {https://arxiv.org/abs/https://www.science.org/doi/pdf/10.1126/sciadv.aaq0194} {https://www.science.org/doi/pdf/10.1126/sciadv.aaq0194} \BibitemShut {NoStop}%
\bibitem [{\citenamefont {Banszerus}\ \emph {et~al.}(2025)\citenamefont {Banszerus}, \citenamefont {Hecker}, \citenamefont {Wang}, \citenamefont {M\"oller}, \citenamefont {Watanabe}, \citenamefont {Taniguchi}, \citenamefont {Burkard}, \citenamefont {Volk},\ and\ \citenamefont {Stampfer}}]{Banszerus2025}%
  \BibitemOpen
  \bibfield  {author} {\bibinfo {author} {\bibfnamefont {L.}~\bibnamefont {Banszerus}}, \bibinfo {author} {\bibfnamefont {K.}~\bibnamefont {Hecker}}, \bibinfo {author} {\bibfnamefont {L.}~\bibnamefont {Wang}}, \bibinfo {author} {\bibfnamefont {S.}~\bibnamefont {M\"oller}}, \bibinfo {author} {\bibfnamefont {K.}~\bibnamefont {Watanabe}}, \bibinfo {author} {\bibfnamefont {T.}~\bibnamefont {Taniguchi}}, \bibinfo {author} {\bibfnamefont {G.}~\bibnamefont {Burkard}}, \bibinfo {author} {\bibfnamefont {C.}~\bibnamefont {Volk}},\ and\ \bibinfo {author} {\bibfnamefont {C.}~\bibnamefont {Stampfer}},\ }\bibfield  {title} {\bibinfo {title} {Phonon-limited valley lifetimes in single-particle bilayer graphene quantum dots},\ }\href {https://doi.org/10.1103/dkgn-pfjb} {\bibfield  {journal} {\bibinfo  {journal} {Phys. Rev. B}\ }\textbf {\bibinfo {volume} {112}},\ \bibinfo {pages} {035409} (\bibinfo {year} {2025})}\BibitemShut {NoStop}%
\bibitem [{\citenamefont {Peres}(2010)}]{Peres2010}%
  \BibitemOpen
  \bibfield  {author} {\bibinfo {author} {\bibfnamefont {N.~M.~R.}\ \bibnamefont {Peres}},\ }\bibfield  {title} {\bibinfo {title} {Colloquium: The transport properties of graphene: An introduction},\ }\href {https://doi.org/10.1103/RevModPhys.82.2673} {\bibfield  {journal} {\bibinfo  {journal} {Rev. Mod. Phys.}\ }\textbf {\bibinfo {volume} {82}},\ \bibinfo {pages} {2673} (\bibinfo {year} {2010})}\BibitemShut {NoStop}%
\end{thebibliography}%

\end{document}